\documentclass[fleqn,usenatbib]{mnras}

\usepackage{newtxtext,newtxmath}
\usepackage{pdflscape} % For landscape pages

\usepackage[T1]{fontenc}

\DeclareRobustCommand{\VAN}[3]{#2}
\let\VANthebibliography\thebibliography
\def\thebibliography{\DeclareRobustCommand{\VAN}[3]{##3}\VANthebibliography}

\usepackage{graphicx}	% Including figure files
\usepackage{amsmath}	% Advanced maths commands
\usepackage[normalem]{ulem}
\title[Forbidden Neon Emission in HAeBe Stars]{Analysis of Forbidden Neon Emission lines in HAeBe Stars using Spitzer IRS spectra}

\author[D. Akhila et al.]{
D. Akhila$^{1}$\thanks{E-mail: akhilakrishnapriya99@gmail.com},
 Blesson Mathew$^{1}$\thanks{E-mail: blesson.mathew@christuniversity.in}, S. Nidhi$^{1, 4}$, B. Shridharan$^{2}$, R. Arun$^{3}$, Hema Anilkumar$^{1}$, G. Maheswar$^{3}$, 
 \newauthor Sreeja S. Kartha$^{1}$, P. Manoj$^{2}$ and Suman Bhattacharyya$^{1}$
\\
$^{1}$Department of Physics and Electronics, CHRIST (Deemed to be University),
Bangalore 560029, India. \\
$^{2}$Department of Astronomy and Astrophysics, Tata Institute of Fundamental Research, Homi Bhabha Road, Colaba, Mumbai 400005, India. \\
$^{3}$Indian Institute of Astrophysics, 2nd Block Koramangala, Bangalore 560034, India.\\
$^{4}$The Oxford College of Science, 17th, 32, 19th Main road, Sector 4, HSR Layout, Bengaluru, Karnataka, 560102, India.}

\date{Accepted XXX. Received YYY; in original form ZZZ}

\pubyear{\the\year{}}

\begin{document}
\label{firstpage}
\pagerange{\pageref{firstpage}--\pageref{lastpage}}
\maketitle

\begin{abstract}
We analyzed high-resolution mid-infrared spectra of 78 well-known Herbig Ae/Be (HAeBe) stars using Spitzer InfraRed Spectrograph data, focusing on the detection of [Ne {\sc II}] and [Ne {\sc III}] emission lines as indicators of ionized outflows or disk winds. Emission from [Ne {\sc II}] at 12.81 $\mu$m or [Ne {\sc III}] at 15.55 $\mu$m was identified in 25 sources, constituting the largest sample of HAeBe stars with these detected lines. Our analysis revealed a higher detection frequency of [Ne {\sc II}] in sources with lower relative accretion luminosity (L$_{acc}$/L$_*$ $<$ 0.1), suggesting a connection to the disk dispersal phase. We examined correlations between neon lines and various spectral features and investigated [Ne {\sc III}]-to-[Ne {\sc II}] line flux ratios to explore potential emission mechanisms. Neon emission is predominantly observed in Group I sources (75\%), where their flared disk geometry likely contributes to the observed emission, potentially originating from the irradiated disk atmosphere. Interestingly, we also find that Group II sources exhibit a higher median relative [Ne\,\textsc{ii}] line luminosity (L$_\mathrm{[Ne\,II]}$/L$_*$), suggesting enhanced photoevaporation rates possibly associated with their more settled disk structures. However, larger samples and higher-resolution spectra are required to confirm this trend definitively. The high detection rate of the [Fe {\sc II}] and [S {\sc III}] lines, commonly associated with EUV-dominated regions, alongside a [Ne {\sc III}]-to-[Ne {\sc II}] emission ratio greater than 0.1 in sources where both lines detected, suggests that EUV radiation is the primary driver of neon emission in our sample.
\end{abstract}


%%%%%%%%%%%%%%%%%%%%%%%%%%%%%%%%%%%%%%%%%%%%%%%%%%

%%%%%%%%%%%%%%%%% BODY OF PAPER %%%%%%%%%%%%%%%%%%

\section{Introduction}

\label{sec:intro}
Herbig Ae/Be (HAeBe) stars are intermediate-mass ($2M_{\odot} \leq M_{*} \leq 8M_{\odot}$) pre-main sequence (PMS) stars \citep{Herbig_1960}. These sources, along with T Tauri Stars (TTS), which are lower-mass analogs, are generally classified as Class II objects in the young stellar object (YSO) evolutionary sequence. Both types of stars exhibit emission lines in the optical and infrared (IR) regions of the spectrum, though they differ in mass and specific evolutionary pathways. These encompass emission lines related to accretion, including the Balmer and Paschen series of hydrogen lines (such as $H\alpha$, $H\beta$, $Pa\beta$, and $Br\gamma$, \citealp{Fairlamb2016}), He I (5876~\AA{} and 6678~\AA{}, \citealp{2001Grinin}), O I (8446~\AA{}, \citealp{1992Hamann,Mathew_2018}), Fe II (4924~\AA{}, 5018~\AA{} and 5169~\AA{}, \citealp{hernandez_2004}) and the Ca II triplet (8498~\AA{}, 8542~\AA{}, 8662~\AA{}, \citealp{2023Ghosh}). Additionally, molecular bands, such as CO, CN, and HCN \citep{2023Stapper,2024Stapper} and silicates \citep{2001Bouwman,2013Sturm}, are detected from the circumstellar disk. As materials from the disk are continuously accreted into the star, they are also channeled along magnetic field lines and some of the materials get ejected from the disk in the form of collimated, high-velocity, ionized jets and broader outflows \citep{2021Ray,2023Pascucci}, resulting in the expulsion of angular momentum and energy acquired from the disk to the surroundings. Prominent atomic and ionic forbidden emission features such as [Ne {\sc II}] (12.81 $\mu$m), [Fe {\sc II}] (24.51$\mu$m, 25.98$\mu$m, 35.34$\mu$m), [S {\sc II}] (6717~\AA{}, 6731~\AA{}), [Si {\sc II}] (34.81$\mu$m), and [O {\sc I}] (6300~\AA{}) arise from jets/outflows and dissociative shocks linked to internal or wind shocks, reflecting the evolving nature of the outflow dynamics \citep{Sperling}. Many forbidden lines are associated with outflows and jets, making their study essential for understanding the physical conditions in these regions.

Neon has a very high ionization potential (21.56 eV), and its role as the dominant species in H {\sc II} regions \citep{Burbidge_1963} makes [Ne {\sc II}] and [Ne {\sc III}] lines particularly significant, serving as excellent tracers of ionizing stars \citep{Ho_2007}. These lines likely originate from gas that is fully ionized by Extreme Ultraviolet (EUV) radiation or partially ionized by X-ray radiation \citep{Pascucci_2014}. They serve as valuable diagnostic tools for probing the gas in the upper layers of the disk, facilitating insights into the interplay between intense stellar high-energy radiation and the disk itself \citep{Liu_2014}. Moreover, the dissipation of the inner disk has been associated with the appearance of a low-velocity component (LVC) in the [Ne {\sc II}] line, with its luminosity increasing as the inner disk dissipates. According to \citet{Pascucci_2020}, this increase is due to the ionization of Ne atoms by hard X-rays, which are no longer obscured by a dense inner disk wind. Therefore, the [Ne {\sc II}] line can serve as a tracer for assessing the timescale of disk dissipation, especially in transition disks with large inner dust holes \citep{2023Pascucci}. In addition to its role as a tracer for disk dissipation, [Ne {\sc II}] emission near young stars can also be explained by three main mechanisms: shocks and jets \citep{Hollenbach_1989,Hollenbach_2009}, irradiated disk atmospheres \citep{Glassgold_2007}, and photoevaporative disk winds \citep{Ercolano_2009,Gorti_2009}.

Most studies on the forbidden emission of neon lines have been conducted on YSOs in general, with a focus on TTS. The initial discovery of the [Ne {\sc II}] 12.81$\mu$m line was made by \citet{Pascucci_2007} in four TTS from a subset of six transition-disk systems, with subsequent detections in TTS reported by \citet{Lahuis_2007} and \citet{Espaillat_2007}. Detections of [Ne {\sc III}] 15.55 $\mu$m line emission from YSOs are much less common, with the earliest reports by \citet{Lahuis_2007}. The most recent detections of neon lines in YSOs have been achieved using James Webb Space Telescope (JWST), as reported by \citet{Bajaj_2024}, \citet{2024_Tychoniec} and \citet{Nisini_2024}. The limited detection of forbidden neon emission lines in HAeBe stars has been largely attributed to the lack for high-energy hard X-rays, which are predominantly observed in their low-mass counterparts \citep{Glassgold_2007}. The first confirmed detection of [Ne {\sc II}] in a Herbig Be star, V892 Tau, was reported by \citet{baldovin_2012}, with the emission arising from photoevaporative wind. There were nine additional reported detections \citep{Salyk_2011, Baldovin_Saavedra_2011, Szulágyi_2012, Sacco_2012, Pascucci_2014, Rigliaco_2015}, having spectral types ranging from B5 to K5, with $\sim$56$\%$ identified as belonging to the `A' spectral type. Focusing on the [Ne {\sc II}] and [Ne {\sc III}] lines in HAeBe stars can provide valuable insights into the high-energy phenomena and emission mechanisms within these stars. The presence or absence of [Ne {\sc II}] emission can reveal important information about the star's evolutionary stage, disk morphology, and the role of photoevaporative winds, helping us better understand the interactions between high-energy radiation and the circumstellar environment.
To study the forbidden neon lines in HAeBe stars, we used the Spitzer Space Telescope \citep{Werner_2004}, which contributed to the study of protoplanetary disks in the mid-infrared (MIR) range \citep{Kim_2016,2010Oliveira,2010Pontoppidan}. This study aims to analyze the MIR (9.9 -- 19.9 $\mu$m and 9.9 -- 37.2 $\mu$m) spectra of 78 well-known HAeBe stars. The analysis utilizes data derived from the collection of Spitzer InfraRed Spectrograph (IRS) spectra for HAeBe stars compiled by \citet{Arun_2023}, which represents the most comprehensive dataset of its kind to date. Examining Ne lines in HAeBe stars using Spitzer spectra lays essential groundwork for future research with more sensitive instruments like the JWST.

This paper is organized as follows: Spitzer survey program and sample selection will be discussed in Section~\ref{sec:data collection}. Section~\ref{sec:analysis} describes the analysis and results on [Ne {\sc II}] and [Ne {\sc III}] line detection, while its implications and potential formation mechanisms are discussed in Section~\ref{sec:Discussion}. Finally, in Section~\ref{sec:conclusions}, we summarize the results.
\section{Data selection}
\label{sec:data collection}
The IRS was one of the key scientific instruments onboard Spitzer space telescope, offering both low (R $\sim$ 60 -- 130) and high-resolution (R $\sim$ 600) spectroscopy in the MIR region (5.2 -- 38 $\mu$m). This instrument has four distinct modules \citep{Houck_2004}, of which we used two that contains both the Ne lines:
 \begin{itemize}
     \item high-resolution, short-wavelength (SH) module covering the range of 9.89 to 19.51 $\mu$m 
     \item  high-resolution, long-wavelength (LH) module for precise observations between 19.83 and 37.14 $\mu$m
 \end{itemize}
 
The data for the study were sourced from the Spitzer Spectral Catalog of HAeBe stars (SSHC), compiled by \citet{Arun_2023}. This catalog contains information on 126 HAeBe stars, among which 94 have low-resolution IRS (SL/LL) spectra, 78 have high-resolution IRS (SH/LH) spectra, and 47 stars have both types of spectra available. Out of the 47 sources with both high- and low-resolution data, 38 showed no detectable lines in the low-resolution spectra. Therefore, the low-resolution data were excluded from the final sample and only the high-resolution spectra from 78 sources were retained for the analysis. Out of the selected 78 sources, 11 have a wavelength range 9.9 -- 19.9 $\mu$m and the remaining 67 stars have a wavelength ranging from 9.9 to 37.2 $\mu$m. It was compiled from the Cornell Atlas of Spitzer/IRS Sources (CASSIS) database, which provides access to both low-resolution and high-resolution Spitzer spectra in staring mode \citep{2011Lebouteiller,Lebouteiller_2015}. CASSIS also offers quantitative assessments of source spatial extent and alternative extraction methods for partially-extended sources, dynamically selecting the best extraction approach based on spatial characteristics. The current versions of the extraction routines for low-resolution and high-resolution spectra are LR7\footnote{https://irs.sirtf.com/Smart/CassisLRPipeline} and HR1\footnote{https://irs.sirtf.com/Smart/CassisHRPipeline}, respectively. High-resolution spectra can be extracted using two methods: optimal extraction and full-aperture extraction \citep{Lebouteiller_2015}. To incorporate circumstellar disk and envelope characteristics for our Class II sources, we used the spectra obtained through full-aperture extraction.

In addition to the MIR spectral data, VLT/X-Shooter spectra and stellar parameters for HAeBe stars were compiled from the literature. Given the established role of X-ray radiation as a driver of [Ne {\sc II}] emission, X-ray data were sourced from various surveys to investigate potential correlations with [Ne {\sc II}] emission in these stars.  This approach enabled a comprehensive examination of the high-energy photons responsible for the emission, offering deeper insight into the circumstellar environments of HAeBe stars.
\section{Analysis and Results}
\label{sec:analysis}
This section outlines the methodologies employed to identify emission lines in the acquired spectra, followed by a detailed comparison of stellar parameters, flux, and luminosity calculations, with results benchmarked against TTS. Additionally, we investigate the potential mechanisms behind neon emission in HAeBe stars by comparing sources with and without neon detection to identify the major excitation processes.

\begin{figure*}
         \centering
         \includegraphics[width=\textwidth]{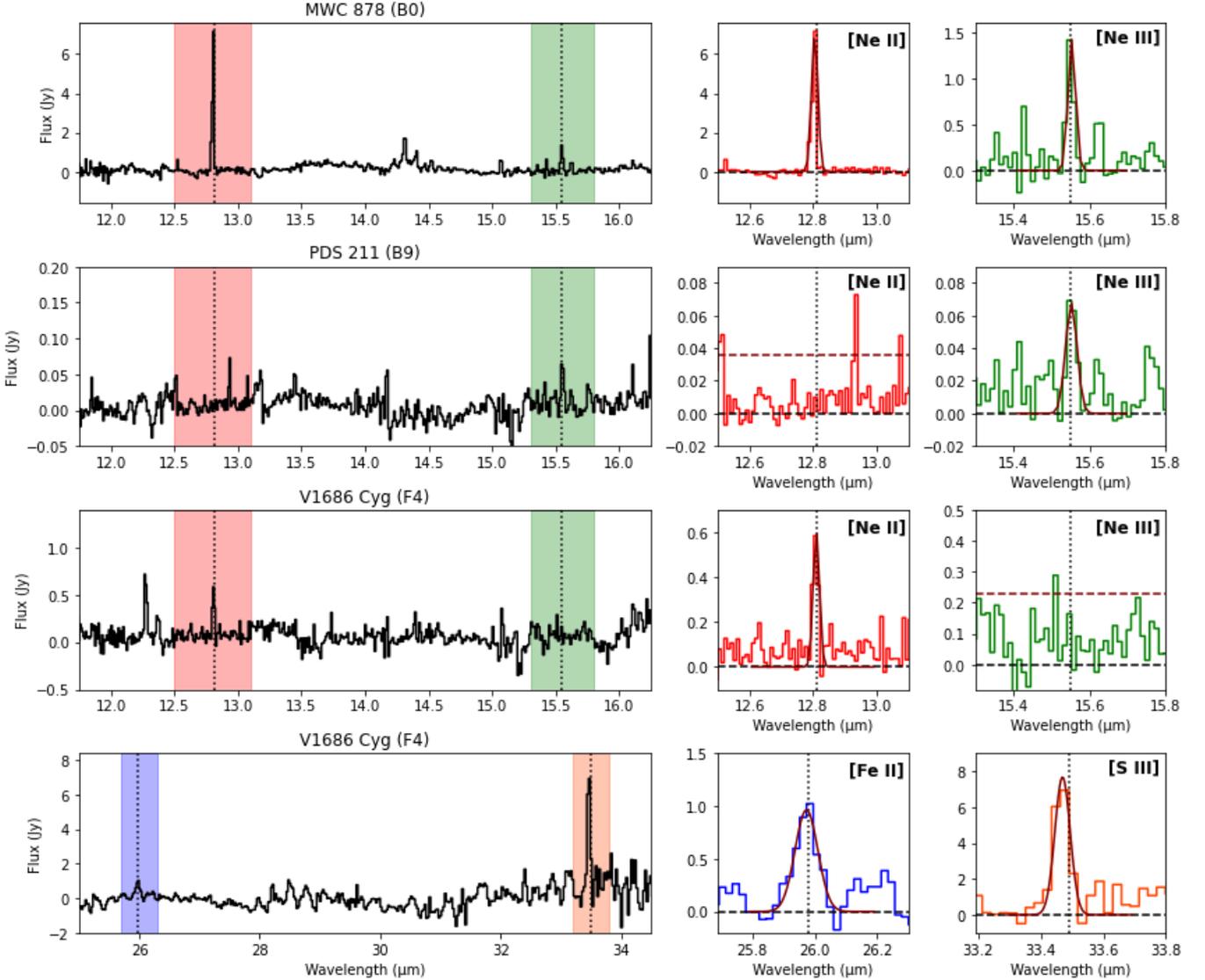}
         \caption{The figure presents the continuum-subtracted spectra for three exemplary sources—MWC 878, PDS 211, and V1686 Cyg—in the left panels, highlighting the [Ne {\sc II}], [Ne {\sc III}], [Fe {\sc II}], and [S {\sc III}] line regions in red, green, blue, and orange, respectively. The corresponding zoomed-in sections of these regions are displayed on the right panels, with detected lines fitted using Gaussian profiles shown in maroon. Non-detections are indicated by 3$\sigma$ thresholds marked with maroon dashed lines. The emission line wavelengths are denoted by black dotted lines in both the full spectra and zoomed-in views, while the baseline in the zoomed-in panels is marked by black dashed lines.}
         \label{fig:spectr_region}
\end{figure*}

\subsection{Neon Line Characterization}
\label{sec: detection}
%\subsubsection{Neon Line Identification Methodology}
Continuum-subtracted, dereddened spectra of 78 sources were obtained using polynomial fitting methods from the \texttt{Pybaselines} package\footnote{https://CRAN.R-project.org/package=baseline}. The dereddening process adopted the extinction law from \citet{1989Cardelli}, utilizing the visual extinction ($A_V$) values taken from \citet{Vioque_2018}. The subsequent line detection involved defining noise regions around the expected line center, with the standard deviation ($\sigma$) calculated from these regions to determine the noise level. The peak flux is defined as the highest value of the baseline-subtracted flux within the specified region. A line was considered detected only when the peak flux exceeded a 3$\sigma$ threshold. 

Given the low resolution (R$\sim$600) of Spitzer, one expects the [Ne {\sc III}] to be blended with the $H_2O$ lines, if present. We followed the criteria outlined by \citet{2010Pontoppidan}, which states that the presence of both $H_2O$ line complexes at 15.17 $\mu$m and 17.22 $\mu$m exceeding 3.5$\sigma$ threshold indicates potential $H_2O$ line contamination of the [Ne {\sc III}] line. After examining our spectra, we found no such detections that exceeded this threshold. This observation is consistent with the expectation of low $H_2O$ abundance in HAeBe star environments, attributed to the high FUV luminosity of these stars, which effectively dissociates $H_2O$ molecules \citep{Fedele_2011,Adams_2019}. Moreover, all sources in our study that display [Ne {\sc III}] detections belong to earlier spectral types (detailed in Section~\ref{sec:Comparison with stellar parameters}). Lower spectral resolution and poor sensitivity challenge the detection of $H_2O$ lines, particularly in Herbig stars. Furthermore, elevated continuum flux and increased noise levels further reduce their detectability \citep{Antonellini_2016}, making detection significantly more difficult.

\begin{landscape}
\setlength\tabcolsep{4.0pt}
\begin{table}
    \centering
    \caption{The table contains the details of RA, DEC, and stellar parameters for 25 HAeBe stars with distinct neon emission lines. Additionally, binarity status, detection of jets/outflows, $n_{(2-25)}$, and X-ray luminosity are included.  }
	\label{tab:multiplicity_table}
      \begin{tabular}{c c c c c c c c c c c c c c c c c c c}
      \hline
        \textbf{Name} & \textbf{RA} & \textbf{DEC} & \textbf{Dist.} & \textbf{T$_{eff}$} & \textbf{Spectral} & \textbf{Meeus} & \textbf{Mass} & \textbf{Age} & \textbf{log($L_*$)} & \textbf{$A_V$} &\textbf{log($\dot{M}_{acc}$)}&\textbf{log($L_{acc}$)}& \textbf{$L_{X}$}  &\textbf{$n_{(2-25)}$}& \textbf{Jets/} & \textbf{System} & \textbf{Ref.} & \textbf{} \\ 
        ~ & (h m s) & (d m s) & (pc) & (K) & \textbf{Type} & \textbf{group} & ($M_{\odot}$) & (Myr) & ($L_{\odot}$) & (mag) &($M_{\odot}$ yr$^{-1}$)& ($L_{\odot}$)& (10$^{30}$erg s$^{-1}$) & ~& \textbf{outflows} & ~ & ~ & ~ \\ \hline
        \textbf{[Ne {\sc II}]} & ~ & ~ & ~ & ~ & ~ & ~ & ~ & ~ & ~& ~ & ~ & ~ & ~ & ~ & ~ & ~ \\ 
        HD 36917  & 05 34 47.0  & -05 34 15  & 474 & 11215 & B8  & II & 3.71 & 0.99 & 2.43 & 0.52&-5.43&1.76&9.12$\pm$0.62$^\complement$&-0.84&... & double lined  & (7) & ~ \\
        & ~ & ~ & ~ & ~ & ~ & ~ & ~ & ~ & ~ & ~ & ~ & ~&~ &~&~& spectral Binary & ~ & ~ \\
        BF Ori &  05 37 13.3  & -06 35 01 & 388.8 & 8970 & A3 & II & 1.81 & 6.38 & 1.29 & 0.33 &-7.03&0.29& 0.32$\pm$0.06$^X$  &-0.63& $\checkmark$& Binary& (10) & ~ \\ 
        HD 38087 & 5 43 0.6 & -2 18 45 & 338.1 & 13600 & B6 & I  & 3.21 & 1.8 & 2.19 & 0.46 &-5.67&1.71& 20.22$\pm$4.44$^\Re$&-0.81& ... & Binary & ... & ~ \\ 
        MWC 137 & 6 18 45.5 & 15 16 52 & 2907.4 & 29000 & B1 & I  & 23.00 & 0.02 & 4.94 & 4.63 &-3.93&3.45& ... &-0.25& $\checkmark$& Close binary & (11) & ~ \\ 
        V590 Mon & 6 40 44.6 & 9 48 2 & 818.4 & 12500 & B7 & ... & 2.30 & 6.00 & 1.38 & 1.03 &...&...&...&0.45 & ... & Binary* & (1),(7) & ~ \\ 
        HD 53367 & 7 4 25.5 & -10 27 16 & 129.7 & 29500 & B0 & ... & ... & ... & 3.13 & 2.05 &...&...& 1.50$\pm$0.78$^X$&-0.93& ... & Visual spectral & (5) & ~ \\ 
        & ~ & ~ & ~ & ~ & ~ & ~ & ~ & ~ & ~ & ~ & ~ & ~ &~&~&~& binary* & ~ & ~ \\
        HD 56895B & 7 18 31.8 & -11 11 34 & 165.3 & 7000 & F3 & II & 1.53 & 8.3 & 0.97 & 0.08 &-6.98&0.14& ... & ... & ... & ... & ... & ~ \\ 
        PDS 37 & 10 10 0.3 & -57 2 7 & 1925.5 & 17500 & B3 & I  & 10.90 & 0.06 & 4.00 & 5.81 &-4.87&2.48& ... &0.07& ... & Massive YSO  & (6),(13) & ~ \\ 
        & ~ & ~ & ~ & ~ & ~ & ~ & ~ & ~ & ~ & ~ & ~ & ~&~ &~&~& binary & ~ & ~ \\
        HD 143006 & 15 58 36.9 & -22 57 16 & 166.1 & 5430 & G9 & I  & 1.56 & 3.70 & 0.46 & 0.31 &-7.5&-0.31& 2.02$\pm$0.55$^\Re$ &-0.61& ... & Binary & (3) & ~ \\ 
        V921 Sco & 16 59 6.79 & 42 42 8.6 & 1545.6 & 29000 & B2 & I  & 20.00 & 0.02 & 4.76 & 4.88 &-4.32&3.15& 109.78$\pm$9.06$^X$ &0.54& $\checkmark$& Binary & (20) & ~ \\ 
        HD 155448 & 17 12 58.8 & -32 14 34 & 953.9 & 10700 & B8  & I  & 4.80 & 0.44 & 2.74 & 0.47 &-5.27&1.88& ... & ... & $\checkmark$&Quintuple system*  & (12) & ~ \\ 
        SAO 185668 & 17 43 55.6 & -22 5 45 & 1481.6 & 16500 & B4 & I & 9.40 & 0.08 & 3.8 & 2.01 &-4.81&2.45& ... &0.21& ... & ... & ... & ~ \\ 
        AS 310 & 18 33 21.2 & 4 58 6 & 2108.4 & 24500 & B0 & I & 11.9 & 0.06 & 4.17 & 4.13 &-4.82&2.79& 10$\pm$5.52$^\complement$&...& $\checkmark$&Binary & (19),(9) & ~ \\ 
        PDS 581 & 19 36 18.9 & 29 32 50 & 687.9 & 24500 & B3 & I & 5.40 & 0.60 & 2.89 & 2.63 &-5.06&2.34& ... &0.22& $\checkmark$& ... & (18) & ~ \\ 
        V1686 Cyg & 20 20 29.3 & 41 21 28 & 1078.8 & 6010 & G0 & I & 2.85 & 1.20 & 1.53 & 1.85 &-6.52&0.51& ...& ... & ... & ... & ... & ~ \\ 
        HD 200775 & 21 1 36.9 & 68 9 48 & 360.8 & 16500 & B4 & I & 5.30 & 0.41 & 3.07 & 1.05 &-5.03&2.24& 115.54$\pm$1.47$^X$&-0.27& ... & Triple system* & (4) & ~ \\ 
        ~ & ~ & ~ & ~ & ~ & ~ & ~ & ~ & ~ & ~ & ~ & ~ & ~ & ~ & ~ & ~ \\ 
        \textbf{[Ne {\sc III}]} & ~ & ~ & ~ & ~ & ~ & ~ & ~ & ~ & ~& ~ & ~ & ~ & ~ & ~ & ~ & ~ \\ 
        PDS 211 & 06 10 17.3 & 29 25 16.6  & 1073.8 & 10700 & B9 & I & 2.41 & 3.00 & 1.79 & 2.98 &...&...& ...& ... & ... & ... & ... & ~ \\ 
        ~ & ~ & ~ & ~ & ~ & ~ & ~ & ~ & ~ & ~ & ~ & ~& ~ & ~ & ~ & ~ & ~ \\ 
        \textbf{[Ne {\sc III}]} \&& \textbf{[Ne {\sc III}]} & ~ & ~ & ~ & ~ & ~ & ~ & ~ & ~ & ~ & ~ & ~ & ~ & ~& ~ & ~ \\ 
        & ~ & ~ & ~ & ~ & ~ & ~ & ~ & ~ & ~ & ~ & ~ & ~& ~ & ~ & ~ & ~ \\ 
        HD 37806 & 5 41 2.3 & -2 43 1 & 427.6 & 10475 & B8 & II & 3.11 & 1.56 & 2.17 & 0.13 &-5.69&1.50& ... &-0.76& ... & Binary* & (2),(15) & ~ \\ 
        HD 259431 & 6 33 5.2 & 10 19 20 & 720.9 & 14000 & B7 & I & 5.2 & 0.42 & 2.97 & 1.11 &-5.25&1.94& 9.54$\pm$1.53$^\complement$& -0.45&$\checkmark$& Binary* & (1),(14) & ~ \\ 
        HD 50138 & 6 51 33.4 & -6 57 59 & 379.9 & 9450 & A1 & I & 4.17 & 0.63 & 2.46 & 0.03 &-4.52&...& ... &-0.43&$\checkmark$& Binary* & (2),(16) \\ 
        PDS 241 & 7 8 38.8 & -4 19 5 & 2887.9 & 26000 & B1 & I & 1.11 & 0.08 & 4.05 & 2.6 &-4.46&2.97& ... &1.79& ... & ... & ... \\ 
        HD 76534 & 8 55 8.7 & -43 28 0 & 910.6 & 19000 & B2 & ... & 7.46 & 0.17 & 3.55 & 0.62 &-4.98&2.43& ... &-1.51& ... & Binary* & (8),(17) \\ 
        HD 130437 & 14 50 50.2 & -60 17 10 & 1653.2 & 24500 & B1 & ... & 13.4 & 0.05 & 4.31 & 2.61 &...&...& 1406$\pm$417$^\Re$&-1.82& ... & ... & ... \\ 
        MWC 878 & 17 24 44.7 & -38 43 51 & 1773.8 & 24500 & B0 & II & 13.5 & 0.05 & 4.32 & 3.06 &-4.54&3.00& ... & -0.27&... & ... & ... \\ 
        PDS 543 & 18 48 0.7 & 2 54 17 & 1413.2 & 29000 & B0 & I & 30.7 & 0.01 & 5.21 & 7.12 &-4.34&3.24& ... &0.16& ... & ... & ... \\ \hline
    \end{tabular}
    \newline
     \raggedright{RA, Dec, Dist., T$_{eff}$, Mass, Age, log($L_*$), and $A_V$ values for all sources are taken from \cite{Vioque_2018}. Meeus Group classifications, log($\dot{M}_{acc}$), and log($L_{acc}$) are from \cite{Guzman-Diaz_2021}. Spectral types are sourced from \cite{Guzman-Diaz_2021}, \cite{Vioque_2018}, and SIMBAD. X-ray data sources are indicated as: `$\complement$' for \textit{Chandra}, `X' for \textit{XMM-Newton}, and `$\Re$' for \textit{eROSITA}. Other parameters are from: Binarity – (1) \citet{Wheelwright_2010}; (2) \citet{Wheelwright_2011}; (3) \citet{Ballabio_2021}; (4) \citet{benisty}; (5) \citet{Pogodin_2006}; (6) \citet{Koumpia_2019}; (7) \citet{Levato_1976}; (8) \citet{Manoj_2002}; (9) \citet{1979Bastian}, Jet/Outflow – (10) \citet{Grinin_2010}; (11) \citet{Mehner_2016}; (12) \citet{Sch_tz_2011}; (13) \citet{Ababakr_2015}; (14) \citet{Li_2014}; (15) \citet{Rucinski_2010}; (16) \citet{Varga_2019}; (17) \citet{Oudmaijer_1999}; (18) \citet{Alcolea_2007}; (19) \citet{1993Goodrich}; (20) \citet{Kraus_2012}. Sources with confirmed binarity are marked with `*'.  }
\end{table}
\end{landscape}
\subsubsection{Comparison with stellar parameters}
\label{sec:Comparison with stellar parameters}
Our analysis using the aforementioned approach detected Ne lines in 25 out of 78 sources (Figure~\ref{fig:neii},~\ref{fig:neii_1} \&~\ref{fig:neiii}).  Detailed information related to the Ne detected sources in each category, including their respective stellar attributes and X-ray emission are provided in Table~\ref{tab:multiplicity_table}. Given that we have a large homogeneous sample of HAeBe stars with neon emission to date, it is imperative to examine the dependence of stellar parameters on the detection of neon lines. The intrinsic and extrinsic stellar attributes are compared for sources with and without Ne line detection within our sample of HAeBe stars. The stellar masses of these sources range from 1.64 to 39.47 $M_\odot$, with ages spanning 0.01 to 8.3 Myr \citep{Vioque_2018}. Additionally, their mass accretion rates ($\dot{M}_{acc}$) fall within 3.16$\times$10$^{-8}$ $M_{\odot} yr^{-1}$ -- 1.17$\times$10$^{-4}$ $M_{\odot} yr^{-1}$ \citep{Guzman-Diaz_2021}. No significant difference in known binarity was observed between sources with and without neon detection, based on the compiled data from the literature \citep{Wheelwright_2010,Pogodin_2006}. Our comparison revealed no clear dependence between the presence of neon emission and several stellar parameters, mirroring the observations commonly seen in the broader context of TTS and YSOs \citep{Flaccomio_2009,Espaillat_2007}. However, a contrasting trend emerged when examining effective temperatures ($T_{eff}$). Within our sample of 78 sources, 19  had $T_{eff}$ $>$ 15000K. Notably, a significantly higher proportion ($\sim$74\%) of these high-$T_{eff}$ sources displayed detectable neon emission. Conversely, for the 59 sources with $T_{eff}$ $\leq$ 15000K, the vast majority ($\sim$81\%) lacked detectable neon emission. Furthermore, while only $\sim$35\% (27 out of 78) of the stellar sample have a mass $>$ 5 $M_\odot$, $\sim$56\% of these massive stars exhibit neon detection. Interestingly, considering all neon-detected sources, a high proportion (60\%) also have a mass $>$ 5 $M_\odot$. From this study, it is identified that Neon lines are preferentially detected in young, massive HAeBe stars, which suggests that high energy radiation from the host star plays a crucial role in the formation of Neon forbidden lines. 

Among these 25 sources with neon detection, 8 exhibited both [Ne {\sc II}] and [Ne {\sc III}] lines, while 16 sources had only [Ne {\sc II}] detection. Interestingly, one source (PDS 211) displayed only [Ne {\sc III}] emission. Representative spectra of three sources showing Neon lines are shown in Figure~\ref{fig:spectr_region}. Upon investigating the distribution of these three groups based on stellar parameters, we observed that the sources exhibiting both [Ne {\sc II}] and [Ne {\sc III}] lines were confined to the spectral range B0 -- A1 (see Figure~\ref{fig:subgrp_psectype}). The distribution of sources with only [Ne {\sc II}] detection was found to be in a broader spectral range, encompassing spectral types B0--G7.  

 \begin{figure*}
         \centering
         \includegraphics[width=\textwidth]{luminosity_spectraltype.pdf}
         \caption{The figure illustrates the variation of $L_{[Ne \;{\sc II}]}$ with spectral types in the upper panel, while the lower panel presents the ratio $L_{[Ne \;{\sc II}]}/L_*$ as a function of spectral type. A total of 41 TTS were included based on the availability of data and spectral parameters compiled from \citet{Baldovin_Saavedra_2011, Gudel_2010, Espaillat_2013, Flaccomio_2009}. Blue circle markers represent sources from our sample, while red triangle markers indicate TTS sources with [Ne {\sc II}] detection. Solid symbols correspond to line detections, and open symbols represent non-detections, with upper limits set at $3\sigma$.}
         \label{fig:spectraltype_luminosity}
\end{figure*}

\subsubsection{Ne line Flux and luminosity measurements}
\label{sec: Ne line Flux and luminosity measurements}
For the 25 sources with Ne line detections, a Gaussian model was fitted to the spectral line using the \texttt{fit\_lines} function from the \texttt{specutils.fitting} module in the \textsc{Specutils} package \citep{nicholas_earl_2023_10016569}. For flux and luminosity calculations, the \texttt{astropy.units} module from the \textsc{Astropy} package \citep{2022Astropy} was utilized to handle units and conversions. The line flux was determined using the Gaussian fit and the \texttt{line\_flux} function from the \texttt{Specutils.analysis} module to integrate flux within specified spectral regions. The derived line fluxes were then converted into luminosities using the distance values from \citet{Vioque_2018}. In instances of non-detections, where the line did not exceed the established threshold for 53 sources, we followed the approach of \citet{Baldovin_Saavedra_2011}, using the $3\sigma$ times the instrumental full width at half maximum (FWHM) as the flux upper limit. From this upper limit, the corresponding upper limit on luminosity was determined. The errors in flux obtained from the fitting, as well as the distance uncertainties, were propagated during the luminosity calculation for sources with detections. The measured [Ne {\sc II}] line luminosity ($L_{[Ne \;{\sc II}]}$) spanned from 6.31$\times$10$^{28}$ erg s$^{-1}$ to 5.02$\times$10$^{33}$ erg s$^{-1}$ for the 24 HAeBe stars (Table~\ref{tab:luminosity and flux values 25}). Furthermore, for sources emitting both lines, $L_{[Ne \;{\sc II}]}$ varied from 1.75$\times$10$^{30}$ erg s$^{-1}$ to 5.02$\times$10$^{33}$ erg s$^{-1}$, whereas [Ne {\sc III}] line luminosities ($L_{[Ne \;{\sc III}]}$) ranged from 7.52$\times$10$^{29}$ erg s$^{-1}$ to 5.53$\times$10$^{32}$ erg s$^{-1}$.

\setlength\tabcolsep{1.85pt}
\begin{table*}
    \centering
    \caption{The table represents the obtained neon line flux ($F_{[Ne \;{\sc II}]}$, $F_{[Ne \;{\sc III}]}$), luminosities ($L_{[Ne \;{\sc II}]}$, $L_{[Ne \;{\sc III}]}$) and the relative luminosities ($L_{[Ne \;{\sc II}]}/L_*$, $L_{[Ne \;{\sc III}]}/L_*$) of the 25 sources with neon detection, along with their Meeus Group classification \citep{Guzman-Diaz_2021}.}
	\label{tab:luminosity and flux values 25}
    \begin{tabular}{cccccccc}
        \hline
        \textbf{Name} &\textbf{Meeus Group}&\textbf{$F_{[Ne \;{\sc II}]}$} & \textbf{$L_{[Ne \;{\sc II}]}$} & \textbf{$F_{[Ne \;{\sc III}]}$} & \textbf{$L_{[Ne \;{\sc III}]}$}&\textbf{$L_{[Ne \;{\sc II}]}/L_*$} &\textbf{$L_{[Ne \;{\sc III}]}/L_*$} \\ 
        ~&~ &(10$^{-14}$erg s$^{-1}$cm$^{-2}$) & (10$^{30}$erg s$^{-1}$) &(10$^{-14}$erg s$^{-1}$cm$^{-2}$) & (10$^{30}$erg s$^{-1}$)& (10$^{-6}$) & (10$^{-6}$)\\ \hline
        HD 36917 &II& 23.5$\pm$0.43 & 6.31$\pm$0.12 & $<$4.93 & $<$1.33 & 4.03$\pm$0.07 & $<$1.28\\ 
        BF Ori &II& 1.53$\pm$0.04 & 0.28$\pm$0.01 & $<$0.69 & $<$0.12 & 5.34$\pm$0.73 & $<$1.65\\ 
        HD 37806 &II& 51.60$\pm$0.64 & 11.29$\pm$0.14 & 23.46$\pm$0.56 & 5.13$\pm$0.60 & 14.72$\pm$0.21 & 6.69$\pm$7.90\\ 
        HD 38087 &I& 1.20$\pm$0.33 & 0.16$\pm$0.04 & $<$0.32 & $<$0.04 & 0.12$\pm$0.03 & $<$0.074\\ 
        PDS 211  &I& $<$1.24 & $<$1.71 & 2.50$\pm$0.19 & 3.45$\pm$0.17 & $<$7.21 & 14.55$\pm$0.89\\ 
        MWC 137 &I& 42.20$\pm$7.07 & 426.87$\pm$36 & $<$2.05 & $<$20.70 & 0.41$\pm$0.03 & $<$0.06\\ 
        HD 259431 &I& 50.82$\pm$0.01 & 31.60$\pm$6.01 & 17.67$\pm$0.25 & 11.06$\pm$0.25 & 10.11$\pm$1.92 & 3.54$\pm$0.08\\ 
        V590 Mon  &...& 16.21$\pm$0.34 & 12.99$\pm$0.28 & $<$2.63 & $<$2.11 & 140.90$\pm$14.63 & $<$22.87\\ 
        HD 50138&I& 204.80$\pm$0.46 & 35.37$\pm$0.46 & 105.64$\pm$1.82 & 18.24$\pm$0.31 & 31.91$\pm$0.46 & 16.45$\pm$0.30\\ 
        HD 53367  &...& 11.73$\pm$2.52 & 0.24$\pm$0.03 & $<$0.39 & $<$0.01 & 0.04$\pm$0.005 & $<$0.001\\ 
        PDS 241 &I& 503.00$\pm$13.09 & 5019.30$\pm$130.96 & 55.45$\pm$0.80 & 553.31$\pm$8.04 & 30.60$\pm$0.80 & 3.37$\pm$0.05\\ 
        HD 56895B &II& 6.51$\pm$0.19 & 0.21$\pm$0.01 & $<$4.52 & $<$0.15 & 5.79$\pm$0.89 & $<$4.12\\ 
        HD 76534 &...& 1.76$\pm$0.40 & 1.75$\pm$0.40 & 0.75$\pm$0.12 & 0.75$\pm$0.12 & 0.085$\pm$0.02 & 0.04$\pm$0.006\\ 
        PDS 37 &I& 13.32$\pm$0.30 & 59.08$\pm$1.35 & $<$5.97 & $<$26.5 &2.38$\pm$0.05 & $<$0.69\\ 
        HD 130437&...& 6.20$\pm$0.15 & 20.27$\pm$0.49 & 1.77$\pm$0.33 & 5.78$\pm$1.09 & 0.26$\pm$0.006 & 0.07$\pm$0.01\\ 
        HD 143006 &I& 1.91$\pm$0.31 & 0.06$\pm$0.01 & $<$1.18 & $<$0.04 & 4.74$\pm$2.16 & $<$3.52\\ 
        V921 Sco  &I& 176.41$\pm$1.47 & 504.25$\pm$4.22 & $<$16.4 & $<$47 & 2.28$\pm$0.02 & $<$0.21 \\ 
        HD 155448 &I& 3.23$\pm$0.04 & 3.52$\pm$0.05 & $<$0.5 & $<$0.54 & 1.38$\pm$0.02 & $<$0.26\\ 
        MWC 878 &II& 243.72$\pm$1.62 & 917.53$\pm$28.8 & 34.47$\pm$3.03 & 129.76 $\pm$0.90 & 4.87$\pm$0.15 & 0.69$\pm$0.004\\ 
        SAO 185668 &I& 4.89$\pm$0.09 & 12.85$\pm$0.26 & $<$1.36 & $<$3.58 & 0.57$\pm$0.01 & $<$0.15\\ 
        AS 310 &I& 420.53$\pm$83.22 & 2236.8$\pm$44.26 & $<$6.15 & $<$32.7 & 26.62$\pm$0.002 & $<$0.57\\ 
        PDS 543 &I& 17.45$\pm$0.24 & 41.70$\pm$0.57 & 3.10$\pm$0.01 & 7.42$\pm$0.23 & 0.09$\pm$0.001 & 0.016$\pm$0.00\\ 
        PDS 581 &I& 33.33$\pm$0.89 & 18.87$\pm$0.50 & $<$10.9 & $<$6.18 & 1.29$\pm$0.03 & $<$2.07\\ 
        V1686 Cyg &I& 21.24$\pm$0.43 & 29.59$\pm$0.59 & $<$6.03 & $<$8.4 & 351.79$\pm$53.07 & $<$64.46\\ 
        HD 200775 &I& 8.76$\pm$0.22 & 1.36$\pm$0.03 & $<$3.38 & $<$0.55 & 0.13$\pm$0.003 & $<$0.12\\ \hline
    \end{tabular}
\end{table*}
\subsubsection{Comparison with TTS}
\label{Comparison with TTS}
\begin{figure}
         \centering
         \includegraphics[width=\columnwidth]{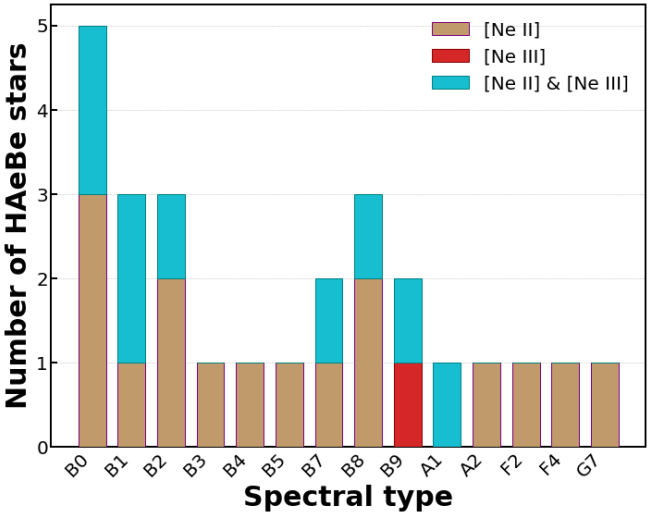}
         \caption{The figure represents the distribution of sources with detected neon emission lines across their spectral types. The brown bars represent sources with only [Ne {\sc II}] emission, the cyan bars represent sources with both [Ne {\sc II}] and [Ne {\sc III}] emission, and the single red bar represents the source with only [Ne {\sc III}] detection.}
         \label{fig:subgrp_psectype}
\end{figure}

The detection rate of [Ne {\sc II}] in Herbig stars is around 32\%, which is lower compared to the 45\% in TTS \citep{Gudel_2010,Espaillat_2013,Baldovin_Saavedra_2011,Flaccomio_2009}. However, this should be taken with caution as the higher continuum flux\footnote{Calculated by integrating extinction-corrected flux over 12.4–12.6 $\mu$m.} in HAeBe stars can mask the [Ne {\sc II}] emission, making it undetectable. The TTS disk models by \citet{Glassgold_2007} (discussed in Section~\ref{sec:Discussion}) predict a $L_{[Ne \;{\sc II}]}$ of $1.4 \times$$10^{28}$ erg s$^{-1}$  and a $L_{[Ne \;{\sc III}]}$ of $1.6 \times$$10^{27}$ erg s$^{-1}$. It was found that the $L_{[Ne \;{\sc II}]}$ values for the HAeBe sources in this study ranged from 10$^{28}$ erg s$^{-1}$ to 10$^{33}$ erg s$^{-1}$. Additionally, the $L_{[Ne \;{\sc III}]}$ values varied from 10$^{29}$ erg s$^{-1}$ to 10$^{32}$ erg s$^{-1}$. For the selected TTS sample, the $L_{[Ne \;{\sc II}]}$ and $L_{[Ne \;{\sc III}]}$ spanned $3\times$10$^{26}$ erg s$^{-1}$ -- $4 \times$10$^{30}$ erg s$^{-1}$ and 7.05$\times$10$^{27}$ erg s$^{-1}$ -- 6.74$\times$10$^{28}$ erg s$^{-1}$, respectively. To investigate the relationship between [Ne {\sc II}] emission and stellar evolution, we examined the $L_{[Ne \;{\sc II}]}$ as a function of spectral type for our HAeBe stars and a compiled sample of TTS with confirmed neon detections from the literature \citep{Gudel_2010,Espaillat_2013,Baldovin_Saavedra_2011,Sacco_2012}. Encompassing the entire spectral range of our HAeBe and TTS sample (B0 to M5), analysis revealed a systematic reduction in the $L_{[Ne \;{\sc II}]}$ as we move from earlier (B0) to later spectral types (M5) in both HAeBe and TTS stars (Upper panel of Figure~\ref{fig:spectraltype_luminosity}). To account for the effect of correlation of stellar luminosity  ($L_*$) with  line luminosity and other parameters, all luminosity parameters were normalized by $L_*$. Interestingly, after normalization, the $L_{[Ne \;{\sc II}]}$/$L_*$ values displayed a similar range from 10$^{-8}$ to 10$^{-4}$, for both TTS and HAeBe stars (Lower panel of Figure~\ref{fig:spectraltype_luminosity}). This uniform distribution across spectral types indicates that the relative luminosity of [Ne {\sc II}] lines does not vary with spectral type. 
\subsection{Investigating Potential Neon Emission Mechanisms } 
\label{sec:all possible mechanisms}
By comparing HAeBe stars with and without neon detection, we explore potential excitation mechanisms responsible for neon emission. Line diagnostics and comparisons with other emission lines will be employed to differentiate between these mechanisms and pinpoint the dominant drivers of neon emission.
\subsubsection{Correlation Between Ne Emission and Accretion}
\begin{figure}
         \centering
         \includegraphics[width=0.9\columnwidth]{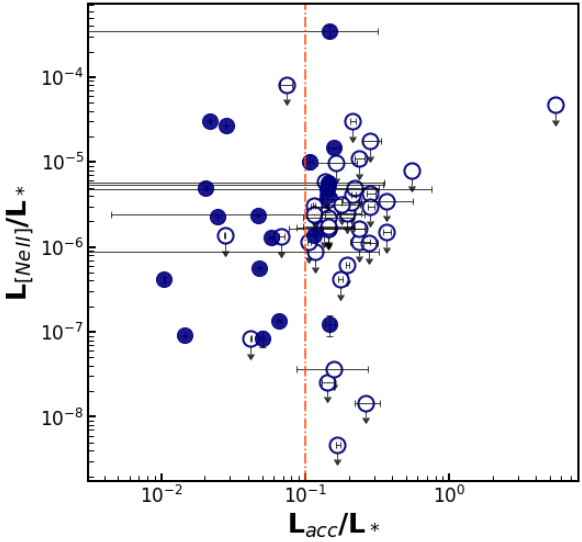}
         \caption{The figure shows the distribution of $L_{[Ne \;{\sc II}]}/L_*$ of HAeBe stars as a function of $L_{acc}/L_*$. Solid and open circle markers denote sources with and without neon detection, respectively, with 3$\sigma$ values plotted as upper limits. The red dash-dot line represents the $L_{acc}$/$L_*$ $=$ 0.1.}
         \label{fig:Lne_lacc}
\end{figure}
To investigate any potential association between Ne line emission and mass accretion, we compared the $\dot{M}_{acc}$ and accretion luminosities ($L_{acc}$) of sources with and without neon detection. The $\dot{M}_{acc}$ values showed no significant difference, ranging from 3.16$\times$10$^{-8}$ $M_{\odot} yr^{-1}$ to 1.17$\times$10$^{-4}$ $M_{\odot} yr^{-1}$ for sources with neon detection and from 5.25$\times$10$^{-8}$ $M_{\odot} yr^{-1}$ to 5.89$\times$10$^{-5}$ $M_{\odot} yr^{-1}$ for non-detections. Interestingly, sources exhibiting both [Ne {\sc II}] and [Ne {\sc III}] lines were confined to a narrower $\dot{M}_{acc}$ range of 10$^{-6}$ $M_{\odot}$ yr$^{-1}$ to 10$^{-4}$ $M_{\odot}$ yr$^{-1}$, contrasting with the broader distribution in sources showing only [Ne {\sc II}]. \citet{Mendigutía2015} observed that $L_{acc}$ correlates with the luminosity of any near-UV, optical, or near-IR emission line, regardless of the physical origin of the spectral transition, possibly due to their common dependence on stellar luminosity ($L_*$) in HAeBe stars. To account for this dependence, $L_{[Ne \;{\sc II}]}$/$L_*$ ratio was plotted against $L_{acc}$/$L_*$, using the $L_{acc}$ obtained from \citet{Guzman-Diaz_2021}. A weak negative correlation was observed between $L_{[Ne \;{\sc II}]}$/$L_*$ and $L_{acc}$/$L_*$ as shown in Figure~\ref{fig:Lne_lacc}, with Pearson's correlation coefficient of -0.33. Moreover, taking into account the observation that there is also no discernible correlation of $\dot{M}_{{acc}}$ with sources with and without [Ne {\sc II}] line \citep{Flaccomio_2009,Espaillat_2007}, the accretion process can be excluded as a potential cause for the formation of [Ne {\sc II}] emission line in our sources. However, it is interesting to observe that the [Ne {\sc II}] detection frequency is higher in sources with lower relative accretion luminosity ($L_{acc}$/$L_*$) $<$ 0.1, which points to a phase of disk dispersal.
 
\subsubsection{Role of X-ray Irradiation in Neon Emission}
\label{sec:Role of X-ray Irradiation in Neon Emission}
High-energy photons, particularly hard X-rays ($\geq$1 keV) are thought to be necessary for exciting neon emission in YSOs \citep{Glassgold_2007,Hollenbach_2009}. We investigate their role in our sample of HAeBe stars. The X-ray luminosity ($L_X$) of HAeBe stars generally lies in the range 10$^{29}$ - 10 $^{33}$ erg s$^{-1}$ \citep[]{2024Anilkumar,1994zinneker}. These detections can be explained by jet collimation shocks \citep[]{2009gutner} or magnetically confined wind shock (MCWS) model \citep[]{1997babel}, potentially driven by primordial magnetic fields inherited from the parent molecular cloud \citep[]{2024Anilkumar}.

\citet{Glassgold_2007} proposed that X-ray irradiation from young stars heats their surrounding disk atmospheres, leading to [Ne {\sc II}] emission within 20 AU of the star. This emission is characterized by a double-peaked line profile with extended wings, potentially indicating high-speed material near the inner disk radius. However, observational studies on the correlation between $L_{[Ne \;{\sc II}]}$ and X-ray luminosity ($L_X$) have produced mixed results. Initial results by \citet{Flaccomio_2009} showed no clear trend between $L_{[Ne \;{\sc II}]}$ and $L_X$, with some [Ne {\sc II}] measurements significantly exceeding predictions, suggesting other factors may influence neon emission. \citet{Lahuis_2007} noted that only about 30\% of sources with [Ne {\sc II}] emission were identified as X-ray sources, implying that X-rays might play a role but may be obscured by incomplete or sensitivity-limited X-ray searches. Later work by \citet{Gudel_2010} found a weak correlation between $L_{[Ne \;{\sc II}]}$ and unabsorbed $L_X$, excluding outliers associated with stellar jets, hinting at a possible link between jet activity and enhanced neon emission. \citet{Sacco_2012} found no clear trend across all samples but observed a weak correlation for optically thick disks, suggesting disk geometry might affect the [Ne {\sc II}] response to X-rays. \cite{Pascucci_2007} reported a positive correlation between $L_{[Ne \;{\sc II}]}$ and $L_X$ using ROSAT data and linear regression methods, establishing a 95\% confidence level despite not correcting for interstellar extinction. The relationship between [Ne {\sc II}] emission and $L_X$ remains complex and lacks a consistent trend across studies. While some research suggests a potential link between X-rays and [Ne {\sc II}] emission, the results are often contradictory. These inconsistencies are likely attributable to factors such as small sample sizes, uncertainties in disk inclination, and variability in $L_X$, highlighting the need for larger, statistically robust samples. In our analysis of HAeBe stars, we addressed these limitations by ensuring the completeness of the dataset, including all sources observed in various X-ray surveys, thus providing a comprehensive foundation for our study.

\begin{figure}
         \centering
         \includegraphics[width=0.9\columnwidth]{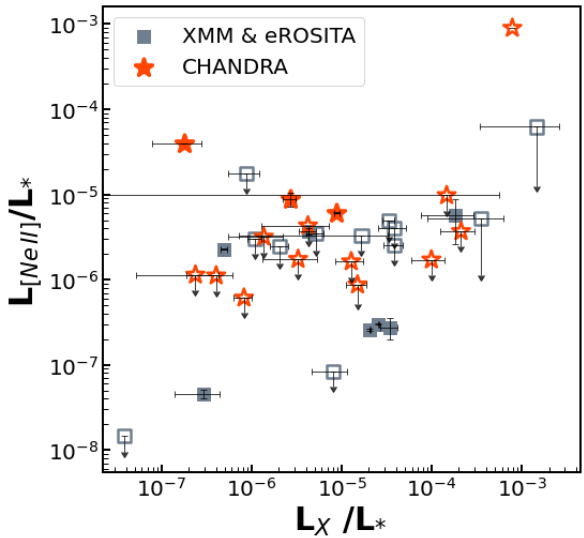}
         \caption{The figure illustrates the distribution of $L_{[Ne \;{\sc II}]}/L_*$ of HAeBe stars as a function of $L_{X}/L_*$. Orange star and grey square markers denote sources for which $L_{X}$ values were taken from \textit{Chandra} and \textit{XMM-Newton} or \textit{eROSITA}, respectively. Solid and open markers represent sources with and without Ne  line detection, respectively, with 3$\sigma$ values plotted as upper limits.}
         \label{fig:lxray_lne}
\end{figure}

To compile a comprehensive X-ray dataset for our sample of 78 HAeBe stars, we leveraged data from three major X-ray space telescopes: \textit{Chandra} \citep{2000Weisskopf}, \textit{eROSITA} \citep{2007Predehl}, and \textit{XMM-Newton} \citep{2000Griffiths}. Prioritizing spatial resolution, we initially checked if our sources were observed by \textit{Chandra} and \textit{XMM-Newton}. Subsequently, we included data from \textit{eROSITA}, a wider-field survey, to ensure completeness. For \textit{Chandra} data, we used the extensive catalog of X-ray emitting HAeBe stars from \citet{2024Anilkumar}. For \textit{XMM-Newton} and \textit{eROSITA}, we crossmatched our sample of 78 HAeBe stars with the XMM-Newton data archive \citep{2019_sarmiento_xmm} and the eROSITA-DE Data Release 1 (eRODat) archive \citep{2024_merloni_erositadr1}. This process revealed that 71 out of the initial 78 sources were covered by these surveys. X-ray detections were confirmed in 47 instances, including 19 with \textit{XMM-Newton}, 15 with \textit{Chandra}, and 13 with \textit{eROSITA}. Notably, 13 of these detections correspond to the same sources observed by multiple instruments. After removing these redundancies, X-ray data are available for 34 unique sources.

Our query of 71 HAeBe stars across multiple X-ray surveys identified X-ray emission in 34 stars. Of these, 10 sources exhibit both X-ray and [Ne {\sc II}] detections. Conversely, 14 stars with [Ne {\sc II}] emission lack X-ray detections, while 24 stars with X-ray detections show no [Ne {\sc II}] emission. In the recent large-scale study on X-ray emission in HAeBe stars by \citet{2024Anilkumar}, 62 were observed with \textit{Chandra}, of which only 44 exhibited X-ray emission. Interestingly, 28 of these sources belonged to the B0-A1 spectral range, similar to the majority of our neon-emitting sources. This suggests a significant association of X-ray emission within this specific spectral class. However, given that nearly all our neon-emitting sources had X-ray observations available, yet only 40\% showed X-ray detection, this statistically indicates that X-rays might not be the primary cause of neon emission in these 25 sources.

To further explore the relationship between X-ray and neon emission, we analyzed the 34 sources with confirmed X-ray detections and derived their X-ray fluxes over the 0.3 – 8.0 keV
range. For sources with \textit{Chandra} data, we used X-ray luminosities from \citet{2024Anilkumar}. X-ray fluxes for sources with \textit{eROSITA} and \textit{XMM-Newton} data were obtained through \textcolor{black}{spectral} model fitting. \textcolor{black}{The details of the data reduction and analysis are provided in Appendix~\ref{appendix:X-ray data reduction and analysis}}. The X-ray luminosities ranged from $8.64 \times 10^{28}$ erg s$^{-1}$ to $1.62 \times 10^{33}$ erg s$^{-1}$, and are provided in Table~\ref{tab:multiplicity_table} and Table~\ref{sec: luminosities 53}. Subsequently, we plotted the $L_{[Ne \;{\sc II}]}/L_*$ as a function of $L_X/L_*$ for these sources, as shown in Figure~\ref{fig:lxray_lne}. We then employed three non-parametric correlation tests: Pearson's, Spearman's rank, and Kendall's rank correlation coefficient. The results, detailed in Table~\ref{tab:correlation_coefficients_macc_xray}, reveal consistently low correlation coefficients and high p-values ($>$ 0.05) across all three tests. These statistical outcomes strongly suggest that X-ray luminosity is not the dominant driver of neon emission in our sample of HAeBe stars.

\begin{figure*}
         \centering
         \includegraphics[width=\textwidth]{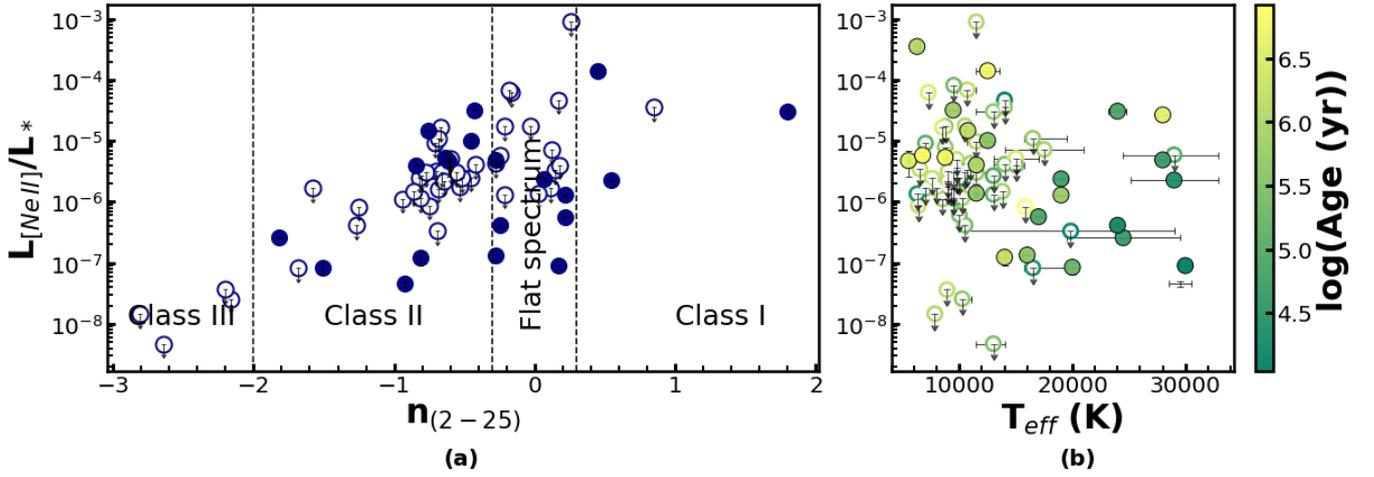}
         \caption{(a) The figure depicts the distribution of $L_{[Ne \;{\sc II}]}/L_*$ of HAeBe stars as a function of n$_{(2-25)}$ index. The solid and open blue symbols represent Ne line detection and non-detection, respectively. The classification of YSOs into Class I, Flat spectrum, Class II, and Class III categories based on the index n$_{(2-25)}$ is depicted by vertical black dashed lines. (b) The figure depicts the distribution of $L_{[Ne \;{\sc II}]}/L_*$ of HAeBe stars as a function of $T_{eff}$, with logarithmic stellar age shown as a gradient. The solid and open circle markers represent sources with and without Ne line detection, respectively, with 3$\sigma$ values plotted as upper limits.}
         \label{fig:lneii_index_teff_age}
\end{figure*}
\begin{table}
    \begin{center}
        
    \caption{The table presents the correlations between the normalized luminosities of mass accretion and X-ray with $L_{[Ne \;{\sc II}]}/L_*$. The corresponding P values are provided in brackets. \label{tab:correlation_coefficients_macc_xray}}
    \begin{tabular}{cccc} 
        \textbf{Lines} & \textbf{Pearson's r} & \textbf{Spearman's $\rho$} & \textbf{Kendall's $\tau$} \\ \hline 
        $L_{acc}/L_*$ & -0.33 & -0.51 & -0.362 \\
        & (0.11) & (0.011) & (0.014) \\
        \hline
        $L_{X}/L_*$ & -0.12 & -0.22 & -0.156 \\
        & (0.74) & (0.53) & (0.60) \\ \hline
    \end{tabular}
    \end{center}
    {Note-Correlations involving [Ne {\sc III}] were not possible due to insufficient data.}
\end{table}
\subsubsection{Role of Disk Morphology and Evolutionary State}
To investigate the contribution of disk morphology to  Ne line emission, we calculated the IR continuum spectral index $n_{(2-25)}$ for the HAeBe stars using the 2MASS $K_s$ magnitude and the ALLWISE \textit{W4} magnitude, following the methods outlined by \citet{1987_lada} and \citet{1989_wilking}. The IR spectral index is used to categorize YSOs as follows: Class I sources have $n_{(2-25)}$$\geq$0.3; flat spectrum sources have 0.3$\geq$$n_{(2-25)}$$\geq$-0.3; Class II objects are characterized by -0.3 $\geq$$n_{(2-25)}$$\geq$-2.0; and Class III sources have $n_{(2-25)}$$\leq$-2.0 \citep{1987_lada,1994_greene,Manoj_2011}. Based on this classification, our sample was found to be inherently dominated by Class II sources. Interestingly, there is an increasing trend in [Ne {\sc II}] strength from Class II to FS to Class I, as shown in Figure~\ref{fig:lneii_index_teff_age}. However, we cannot confirm the result of \cite{Flaccomio_2009}, which suggests stronger [Ne {\sc II}] emission in Class I sources compared to Class II, due to the limited number of Class I Herbig stars in our sample.

 High-velocity outflows from young stars can generate shocks that heat the surrounding gas, creating environments conducive to [Ne {\sc II}] emission, particularly at velocities exceeding 40–50 km/s \citep{Hollenbach_1989, Hollenbach_2009, Shang_2010}. Consistent with this, \citet{Gudel_2010} found that Class II sources exhibiting jets or outflows showed significantly stronger [Ne {\sc II}] emission, with line luminosities 1–-2 orders of magnitude higher than sources without outflows \citep{Rab_2016}. Our results support this trend, with median $L_{[Ne \;{\sc II}]}$ values of $3.23 \times 10^{31}$ erg s$^{-1}$  for sources with outflows compared to $5.25 \times 10^{30}$ erg s$^{-1}$ for those without. Furthermore, the notably high $L_{[Ne \;{\sc II}]}$ in MWC 137 and AS 310 can likely be attributed to the presence of jets or outflows from the source \citep{Mehner_2016, 1993Goodrich}. These findings underscore a possible correlation between [Ne {\sc II}] emission and the presence of jets or outflows in YSOs, although definitive conclusions regarding their role in [Ne {\sc II}] emission from HAeBe stars remain elusive and require further investigation.

 The classification introduced by \citet{meeus2001} divides YSOs into Group I and Group II based on the shape of their infrared spectral energy distributions (SEDs). Group I sources exhibit a rising MIR excess, modeled by a combination of a power-law and blackbody component, indicative of flared outer disks. In contrast, Group II sources display a declining MIR excess, fitted by a power-law only, suggesting more settled, compact disks without flaring. Cross-matching our sample with the classifications from \citet{Guzman-Diaz_2021}, we identified Meeus classifications for 60 out of 78 sources.

Upon analyzing the distribution of sources based on Meeus classification, we find that 60 out of 78 sources have a defined classification, with 32 classified as Group I and 28 as Group II. Among these, [Ne \textsc{ii}] emission is detected in 15 Group I sources (47\%) and 5 Group II sources (18\%). Of the 25 sources in the full sample with detected [Ne \textsc{ii}] emission, 20 have a Meeus classification, of which 75\% (15 out of 20) belong to Group I. This suggests that Ne emission is more prominent in these younger, rapidly evolving sources, which often harbor flared disks or disks with inner holes formed due to photoevaporation \citep{meeus2001}. Furthermore, most of the Neon-detected sources exhibit high MIR luminosity, low relative accretion luminosity, and belong to Group I, supporting the flared disk morphology. The flared structure enhances direct ionization of Ne atoms by stellar UV radiation, though the resulting Neon gas may or may not be bound to the disk. Additionally, in disks with inner holes, these gaps can facilitate the penetration of higher-energy radiation from the central star to the outer disk regions \citep{Pascucci_2020}. Following the depletion of inner disk material due to photoevaporation, the exposed inner rim becomes directly irradiated by the stellar radiation field \citep{2006Alexander}.  Hence, the observed Ne emission may originate from the irradiated disk atmosphere, where Neon atoms are ionized by stellar radiation. 

The ratio of $L_{[Ne \;{\sc II}]}/L_*$ for 78 sources was analyzed in relation to mass, $T_{eff}$, and disk classification. A decreasing trend in $L_{[Ne \;{\sc II}]}/L_*$ was observed with increasing mass and $T_{eff}$. Notably, older sources exhibited higher $L_{[Ne \;{\sc II}]}/L_*$ values compared to younger counterparts as seen in Figure~\ref{fig:lneii_index_teff_age}.  Sources with higher $L_{[Ne \;{\sc II}]}/L_*$ ratios were predominantly found in the Meeus Group II classification (Table~\ref{tab:multiplicity_table}).

To quantify the trend in [Ne\,II] emission among the Meeus groups, we computed both the median relative line luminosities (\( L[\mathrm{Ne\,II}]/L_* \)) and absolute [Ne\,II] line luminosities. The median relative luminosity for Group II sources is \( 5.35 \times 10^{-6} \), which is higher than that of Group I sources (\( 2.33 \times 10^{-6} \)). However, in terms of absolute [Ne\,II] line luminosities, Group I sources show stronger emission, with a median value of \( 3.06 \times 10^{31}~\mathrm{erg\,s^{-1}} \), compared to \( 6.32 \times 10^{30}~\mathrm{erg\,s^{-1}} \) for Group II sources. These values indicate that while Group II sources tend to have stronger [Ne\,II] emission relative to their stellar luminosity, Group I sources are intrinsically brighter in [Ne\,II].

Although [Ne {\sc II}] emission was detected in fewer Group II sources compared to Group I, those that exhibited emission had a significantly higher median of relative line luminosity. This observed trend is likely associated with the settled disk structure of Group II sources. In these systems, the settling of dust might reduce the overall interception of high-energy radiation by the disk compared to flared disks, allowing ionizing photons to penetrate deeper into the disk. This would facilitate a more efficient penetration of these photons to the outer regions of the disk, potentially accelerating the photoevaporation process \citep{Gorti_2009}, and contributing to the disk dispersal. Alternatively, in self-shadowed Group II disks, the outer region could be shielded from the central star's radiation by the inner disk. This could lead to a scenario where high-energy photons preferentially interact with the exposed inner wall of the disk \citep{2004Dullemond}, potentially stimulating neon emission from the irradiated disk atmosphere. The observed trend of comparatively higher $L_{[Ne \;{\sc II}]}/L_*$ ratios in older, less massive Group II HAeBe stars aligns with the possibility of an enhanced photoevaporation rate in these systems, where the settled disk configuration could play a role.

Our sample primarily consists of Class II and flat-spectrum sources, limiting direct comparison of line strengths between Class I and Class II objects. The significant presence of neon-emitting sources in Meeus Group I suggests a potential link to disk features like flared structures or inner holes. Higher $L_{[Ne \;{\sc II}]}/L_*$ ratios in older, less massive Group II stars may indicate enhanced photoevaporation in their settled disk structures. However, current evidence is insufficient to establish whether Group II sources are systematically older than Group I. High-resolution MIR observations using instruments such as VISIR are essential to confirm these trends, investigate the kinematics of Neon emission, and clarify the role of photoevaporation in compact Group II disks.
\subsubsection{ Correlation analysis with other forbidden lines}
\begin{table}
    \begin{center}
    \caption{The table shows the detection rates of forbidden lines in HAeBe stars.The number of sources with line detection, out of the total number of spectra considered, is indicated in brackets under the detection rates.}
    \begin{tabular}{ccc}
        \textbf{Line} & \textbf{ Ne detection}& \textbf{ Ne non-detection}\\ \hline  
        [Fe {\sc II}] & 68\% & 28.3\% \\ 
        (25.99\micron)&(17/25)&(15/53)\\
        \hline
        [S {\sc III}] & 76\% & 9.4\% \\ 
        (33.49\micron)&(19/25)&(5/53)\\
        \hline
        [O {\sc I}] & 60\% & 66\% \\ 
        (6300~\AA{})&(6/10)&(19/29)\\
         \hline
    \end{tabular}
    \end{center}
    \label{tab:detection rate}
\end{table}
\begin{figure*}
         \centering
         \includegraphics[width=\textwidth]{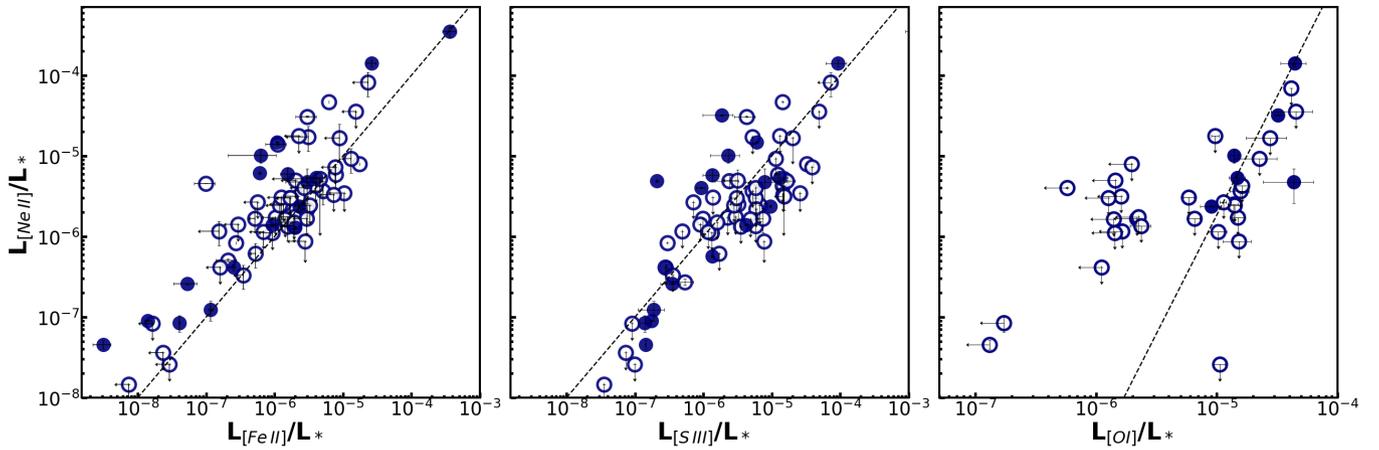}
         \caption{The figure shows the distribution of $L_{[Ne \;{\sc II}]}/L_*$ of HAeBe stars as a function of $L_{[Fe \;{\sc II}]}$/$L_*$, $L_{[S \;{\sc III}]}$/$L_*$ and $L_{[O \;{\sc I}]}$/$L_*$, with the black dashed line representing the one-to-one correlation for comparison. The solid and open circle markers represent sources with and without  Ne detection, respectively, with 3$\sigma$ values plotted as upper limits.}
         \label{fig:lneii_lfeii_lsiii}
\end{figure*}

To explore the possible regions and mechanisms of  Ne line formation, we investigated the presence of various other lines, particularly forbidden lines in the MIR and optical ranges. This analysis utilized X-Shooter and Spitzer data, examining spectra of sources both with and without  Ne line detection. X-Shooter, a multi-wavelength (3000 -- 25000~\AA{}) spectrograph, provided spectra for 44 of the initial 78 sources. For the analysis, we considered the wavelength range of 5595 -- 10240~\AA{} covered by the Visual arm of X-Shooter for 39 spectra with high SNR \citep{2011Vernet}.

The [O {\sc I}] line at 6300~\AA{} can originate from disk winds, jets, or bound disk layers in YSOs \citep{Hartigan_1995,Acke_2005}. Studies by \citet{2011Pascucci} and \citet{baldovin_2012} found that [O {\sc I}] and [Ne {\sc II}] generally arise from different regions, while \cite{Gudel_2010} suggested a possible disk origin for [Ne {\sc II}]. These findings highlight the need for further comparisons to clarify the conditions and mechanisms driving [Ne {\sc II}] and [O {\sc I}] emissions. Using a 3$\sigma$-based detection criterion, we find that [O\,\textsc{I}] emission is present in $\sim$60\% of sources with [Ne\,\textsc{II}] emission and $\sim$66\% of sources without it, suggesting a relatively comparable detection rate across both groups, irrespective of [Ne {\sc II}] emission. Interestingly, we observe that the majority of sources lacking both [Ne\,\textsc{II}] and [O\,\textsc{I}] emission belong to Meeus Group II (8 out of 9), while those exhibiting both lines are primarily associated with Group I (5 out of 6). This trend suggests that disk geometry and vertical structure play a key role in shaping the presence of these gas tracers.

We estimated the $L_{[O \;{\sc I}]}$ by calculating equivalent widths from the continuum-normalized X-Shooter optical spectra. Johnson's R-band fluxes, derived from an extinction-corrected theoretical spectra using BT-NextGen model \citep{2006Barber,2009Asplund,2011Allard,2012Allard} with the VO SED Analysis tool (VOSA\footnote{http://svo2.cab.inta-csic.es/theory/vosa/}; \citealt{2008Bayo}), were adopted as continuum fluxes to ensure homogeneity, as observed R-band magnitudes were unavailable for most sources. Line fluxes were then computed as products of these continuum fluxes and line equivalent widths \citep{1990Waller}. Error estimations followed the methods established in Section~\ref{sec: Ne line Flux and luminosity measurements}. For non-detections, luminosity upper limits were estimated using upper limits on the equivalent width, calculated as three times the product of the mean standard deviation of the continuum and the instrumental FWHM, where FWHM is given by $\lambda$/R, with $\lambda$=6300\,\AA\ and R representing the spectral resolution. The $L_{[O \;{\sc I}]}$/$L_*$ and $L_{[Ne \;{\sc II}]}$/$L_*$ (Figure~\ref{fig:lneii_lfeii_lsiii}) exhibit a moderate positive correlation (Table~\ref{tab:correlation_coefficients}), suggesting a shared emission region. While both lines serve as tracers of circumstellar material in YSOs, their distinct emission properties offer complementary insights. The [O {\sc I}] line, sensitive to lower-energy photons, typically probes cooler, denser gas, whereas the [Ne {\sc II}] line, sensitive to higher-energy photons, can originate from both hot, ionized gas and cooler, partially ionized regions \citep[figure 5]{2023Pascucci}. This broader range of excitation conditions for [Ne {\sc II}] suggests that it may trace a wider array of physical environments, including the upper layers of a photoevaporative wind at larger radii, as proposed by \cite{Rigliaco_2013}, while [O {\sc I}] may probe either bound disk gas or lower layers of the same wind. A key limitation is the inability to resolve the [O {\sc I}] line into HVC and LVC components, preventing precise differentiation between jets, disk winds, or bound disk layers, highlighting the need for higher spectral resolution to constrain the line origin. Furthermore, while the observed correlation supports a common origin, this trend is subject to statistical constraints due to the small number of sources with robust detections of both lines.

Ne-emitting sources displayed higher detection rates for other MIR lines that trace high-energy, including [Fe {\sc II}] (25.99$\micron$) and [S {\sc III}] (33.49$\micron$). Specifically, out of 25  Ne sources, 20 and 17 sources had [Fe {\sc II}] and [S {\sc III}] detections, respectively (Table~\ref{tab:detection rate}). The sample spectra for a source over the wavelength region of [Fe {\sc II}] and [S {\sc III}] emission lines are shown in Figure~\ref{fig:spectr_region}. The [Fe {\sc II}] lines at 17.94$\micron$ and 25.99$\micron$, which are doublets with lower ionization potentials than hydrogen, are typically observed in ionized and warm neutral gas, indicating high-energy radiation, atomic shocks, or jets/outflows \citep{Hollenbach_1989}. In our sample, the detection rate of the 17.94 $\micron$ line was notably low, likely due to its luminosity being roughly one-third of that of the 25.99 $\micron$ line, resulting in a negligible value in our analysis. Moreover, the similarity in charge exchange rates between [Ne {\sc III}] and [S {\sc III}] with HI \citep{1980butler} suggests that [Ne {\sc III}] would likely be more prevalent in regions dominated by EUV radiation \citep{Espaillat_2013}. 

\begin{table}
    \begin{center}        
    \caption{The table depicts the correlations between the normalized luminosities of [Fe {\sc II}], [S {\sc III}] and [O {\sc I}] with $L_{[Ne \;{\sc II}]}/L_*$. The corresponding P values are provided in brackets.}
    \label{tab:correlation_coefficients}
    \begin{tabular}{cccc} 
        \textbf{Lines} & \textbf{Pearson's r} & \textbf{Spearman's $\rho$} & \textbf{Kendall's $\tau$} \\ \hline 
        $L_{[Fe \;{\sc II}]}/L_*$ & 0.95 & 0.78 & 0.647 \\ 
        & (0.000) & (0.0002) & (0.0001) \\ \hline
        $L_{[S \;{\sc III}]}/L_*$ & 0.95 & 0.78 & 0.637 \\ 
        & (0.000) & (0.0004) & (0.0002) \\ \hline
        $L_{[O \;{\sc I}]}/L_*$ & 0.61 & 0.60 & 0.47 \\ 
        & (0.20) & (0.21) & (0.27) \\ \hline
    \end{tabular}
    \end{center}
    
    {Note-Correlations involving [Ne {\sc III}] were not possible due to insufficient data.}  
\end{table}
 
By utilizing the luminosities of [Fe {\sc II}] ($L_{[Fe \;{\sc II}]}$), [S {\sc III}] ($L_{[S \;{\sc III}]}$) and $L_*$, the ratios $L_{[Ne \;{\sc II}]}$/$L_*$ were plotted against $L_{[Fe \;{\sc II}]}$/$L_*$ and $L_{[S \;{\sc III}]}$/$L_*$ to explore potential correlations between the lines (Figure~\ref{fig:lneii_lfeii_lsiii}). Correlation analyses were not performed for [Ne {\sc III}] due to the limited number of sources with detections of both [Ne {\sc III}] and the [Fe {\sc II}] or [S {\sc III}] lines. This low sample size would render the correlation coefficients statistically unreliable. A strong positive correlation has been identified between [Ne {\sc II}], [Fe {\sc II}] and [S {\sc III}], evidenced by the statistical coefficients given in Table~\ref{tab:correlation_coefficients}. These results suggest a potential shared emission region for [Ne {\sc II}], [Fe {\sc II}], and [S {\sc III}]. Furthermore, the detection of [Fe {\sc II}] and [Ne {\sc II}] in the collimated jets of low-luminosity protostars using JWST by \citet{Narang_2024} raises the possibility of similar phenomena in HAeBe stars. 

\subsubsection{Neon Line Ratios as Probes of Disk Ionization Mechanisms}
Neon emission lines serve as valuable diagnostics for differentiating high-energy processes in YSOs, such as ionization by X-rays, EUV radiation, and shocks \citep{1987Hartigan,Hollenbach_2009,Glassgold_2007}. A detailed discussion of these mechanisms and their implications is presented in Section~\ref{sec:Discussion}. The [Ne {\sc III}]-to-[Ne {\sc II}] line flux ratio is particularly sensitive to the dominant excitation source, as each high-energy process uniquely affects disk ionization and photoevaporation \citep{Espaillat_2023}. The observed ratio in YSOs can be attributed to the varying ionization and penetration effects of different photons—soft and hard X-rays, as well as soft and hard EUV radiation \citep{Glassgold_2007,Meijerink_2008,Hollenbach_2009}. In soft X-ray-dominated regions, the ratio typically falls below 0.1, as neon ionization is primarily driven by secondary electrons, resulting from high absorption of X-rays by heavier elements like He, C, and O. Additionally, the gas is hotter due to the higher cross-sections for softer X-rays, causing more heating per unit volume \citep{Hollenbach_2009}. In these X-ray-dominated layers, the prevalence of neutral hydrogen facilitates efficient charge exchange, converting [Ne {\sc III}] to [Ne {\sc II}] and leading to larger [Ne {\sc II}] luminosities relative to [Ne {\sc III}] \citep{Glassgold_2007}. In hard X-ray ($>$1 keV) regions, direct photoionization of neon creates higher ionization states, increasing the [Ne {\sc III}]-to-[Ne {\sc II}] ratio. In soft EUV-dominated regions, the limited penetration restricts ions to regions with substantial neutral hydrogen, where charge exchange further lowers the [Ne {\sc III}] abundance by converting it to [Ne {\sc II}]. As a result, in both these cases, the partially ionized gas reduces the hydrogen abundance, diminishing the conversion of [Ne {\sc III}] to [Ne {\sc II}], and yielding ratios greater than 0.1. In contrast, hard EUV-dominated regions are fully ionized with low hydrogen abundance, which reduces the efficiency of charge exchange. Thus, in EUV-irradiated layers, [Ne {\sc III}] is more abundant, potentially leading to $L_{[Ne \;{\sc III}]}$ exceeding $L_{[Ne \;{\sc II}]}$ \citep{1980butler,Hollenbach_2009}, resulting in a ratio greater than 1. This characteristic behavior in neon ionization allows us to link observed line ratios to specific excitation mechanisms, providing insight into the photoevaporation efficiency driven by various high-energy sources.

\begin{figure}
         \centering
         \includegraphics[width=0.9\columnwidth]{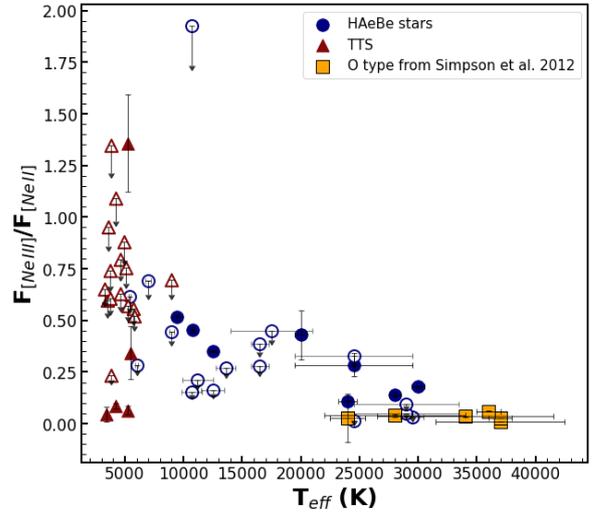}
         \caption{The figure shows the distribution of [Ne {\sc III}]-to-[Ne {\sc II}] line flux ratio as a function of $T_{eff}$. The blue circle and red triangle markers represent HAeBe stars and TTS, respectively. The yellow square markers represent the sample of O-type YSOs from \citet{Simpson_2011}. Solid symbols indicate detection of both lines in the pair, while open symbols signify non-detection of  Ne line, with 3$\sigma$ value plotted for the upper limit.}
         \label{fig:tts_haebe_ratio_teff}
\end{figure}

In our dataset of 25 sources, 8 emit both [Ne {\sc III}] and [Ne {\sc II}], with all showing [Ne {\sc III}]-to-[Ne {\sc II}] line flux ratios below 1, indicating that the dominant photons were hard/soft X-rays or soft EUV \citep{Hollenbach_2009}. We conducted a comparison of flux ratios with other spectral parameters, including confirmed multiplicity, variability, as well as jet and outflow characteristics. Notably, HD 50138, which exhibits jets/outflows \citep{Oudmaijer_1999}, has the highest ratio, around 0.52. A total of 20 TTS with both [Ne {\sc II}] and [Ne {\sc III}] flux values available \citep{Lahuis_2007, Najita_2010} were considered for the line flux ratio analysis with the HAeBe sample. Among these, 5 displayed detection for both lines, while the remainder only showed [Ne {\sc II}] emission. We combined flux data from literature sources with those in our sample to calculate the [Ne {\sc III}] to [Ne {\sc II}] flux ratios. For sources where only one line was detected, we utilized 3$\sigma$ values as upper limits for the calculation. 

Among the TTS sources analyzed, one displayed an exceptionally high [Ne {\sc III}]-to-[Ne {\sc II}] ratio, the highest in the sample of TTS and Herbig stars. This is likely due to the low $\dot{M}_{acc}$ ($<$10$^{-8}$ $M_{\odot} yr^{-1}$) and mass loss rate ($<$ 10$^{-9}$ $M_{\odot} yr^{-1}$), facilitating the penetration of EUV radiation into the stellar wind and subsequent interaction with the disk \citep{Espaillat_2013}. Additionally, three sources with detection of both lines exhibited a ratio $<$ 0.1, suggesting emission dominated by hard X-rays. For Herbig stars, the observed [Ne {\sc III}]-to-[Ne {\sc II}] ratios, typically between 0.1 and 1, as shown in Figure~\ref{fig:tts_haebe_ratio_teff}, indicate that the dominant radiation source is either hard X-rays or soft EUV. However, our analysis shows no significant correlation with X-rays (Section~\ref{sec:Role of X-ray Irradiation in Neon Emission}). Given the significant EUV flux from these stars and the high occurrence of the [S {\sc III}] line, which is typically associated with EUV-dominated regions, the intense EUV radiation from the central source is likely the primary driver of  Ne line formation \citep{Alexander_2008}.

We examined the relationship between stellar mass, T$_{\rm eff}$ and the mechanisms driving neon emission in TTS and Herbig stars, with a focus on the contributions from high-energy photons such as X-rays and EUV radiation. The extended magnetospheres of low-mass TTSs enhance the production of high-energy photons, which in turn excite neon ions and lead to increased neon emission \citep{Hussain_2009, 2012Gregory}. However, this effect diminishes with increasing stellar mass, potentially limiting the detectability of neon emission in more massive stars, as illustrated in Figure~\ref{fig:spectraltype_luminosity}. For intermediate-mass HAeBe stars, neon line emission in the outer disk is likely driven by either hard/soft X-rays or soft EUV radiation. In contrast, for high-mass Herbig stars with significant EUV luminosities, [Ne {\sc III}] emission can saturate due to high electron densities \citep{Hollenbach_2009}. This saturation effect results in a decrease in the [Ne {\sc III}]-to-[Ne {\sc II}] line ratio with increasing T$_{\rm eff}$. A similar trend was observed in O-type YSOs by \citet{Simpson_2011}, where the [Ne {\sc III}]-to-[Ne {\sc II}] ratio ranged from 0.008 to 0.059 (Figure~\ref{fig:tts_haebe_ratio_teff}), which is lower than the values observed in our analysis, further supporting the idea that saturation effects are more pronounced in higher-mass stars.

To assess the impact of shocks on neon emission in Herbig stars, a comparison with theoretical shock models is essential. For shocks with velocities between 30--40 km s$^{-1}$, models by \citet{Hollenbach_1989} predict that [Fe {\sc II}] 26 $\mu$m and H$_2$ S(1), H$_2$ S(2), H$_2$ S(3) emission should dominate, with these lines being up to 1-3 orders of magnitude stronger than [Ne {\sc II}]. However, our data show the opposite trend: [Ne {\sc II}] emission consistently exceeds these lines by 1–2 orders of magnitude. Additionally, our data show limited H$_2$ emission, with H$_2$ S(1) at 17.03 $\mu$m detected in only $\sim$18\% of the total sample (9 with neon emission) and H$_2$ S(0) at 28.5 $\mu$m in $\sim$30\% of sources (10 with neon detection). This low level of H$_2$ emission further suggests that low-velocity shocks are unlikely to be the dominant mechanism driving [Ne {\sc II}] emission \citep{1996Kaufman}. For high velocity shocks, models predict the detection of [Fe {\sc I}] 24 $\mu$m and [S {\sc I}] 25.3 $\mu$m, with [S {\sc I}] typically being stronger \citep{Hollenbach_1989}. In our sample, however, [Fe {\sc I}] and [S {\sc I}] lines were detected in only one source each, suggesting that strong shocks are not the reason for [Ne {\sc II}] in HAeBe stars. Furthermore, our observed H I (7-6) to [Ne {\sc II}] line ratios range from 0.04 to 2, with one exception (HD 200775), which shows a ratio of 4.07, well above the theoretical value of 0.008 expected for EUV- and X-ray-illuminated shocks \citep{Hollenbach_2009}. This discrepancy suggests that the origin of the observed H I emission likely arises from regions with higher densities than the critical density of [Ne {\sc II}]. The available data supports EUV photoionization as the dominant driver of [Ne\,{\sc II}] emission in HAeBe stars, although a contribution due to outflows cannot be completely ruled out given the correlation of higher [Ne\,{\sc II}] emission in sources known to drive jets and outflows.
\section{Discussion}
\label{sec:Discussion}
The major mechanisms of neon ionization in YSOs include the ejection of electrons from the outer (L) and inner (K) shells by EUV or X-ray photons. The resulting ionization state depends on the X-ray energy and K-shell thresholds (as suggested by \cite{Glassgold_2007}). Interestingly, experiments show the formation of higher ions like $Ne^{+3}$ and $Ne^{+4}$ from K-shell vacancies in neutral Ne, which then revert to $Ne^{+2}$ through charge exchange with atomic hydrogen in protoplanetary disk atmospheres \citep{carlson_1965,Krause_1964}. This proposed mechanism was based on a thermal-chemical model for a generic T Tauri disk by \citet{1999D'Alessio}, exposed to a strong stellar X-ray flux. Subsequent detections of MIR forbidden Ne lines in low-mass YSOs, particularly TTS, have been reported, with limited studies investigating the underlying mechanisms \citep{Pascucci_2007,baldovin_2012}. Energetic photons originating from both accretion hotspots and chromospheric activity can influence neon emission \citep{Alexander_2005}. However, the effectiveness of these photons depends on the accretion rate. Crucially, low $\dot{M}_{acc}$ ($<$ $10^{-8}$ $M_{\odot} yr^{-1}$) allow EUV radiation to penetrate the circumstellar disk wind \citep{1998hartmann}, amplifying Ne emission beyond what is produced solely by X-ray irradiation. Additionally, \citet{Alexander_2008} proposed that the production of $L_{[Ne \;{\sc II}]}$ could be explained by a photoevaporative wind model, with the UV flux from the central star as the ionizing source. These models provide a framework for understanding the observed neon emission in YSOs. To test these predictions and gain insights into the dominant ionization mechanisms, we conducted a comparative analysis of neon emission lines in TTS and HAeBe stars.

A comparative analysis of neon emission lines in YSOs reveals a significantly higher frequency of [Ne {\sc II}] detections in TTS compared to HAeBe stars (Section~\ref{Comparison with TTS}). This disparity is primarily due to the abundance of high-energy X-rays in the low-mass TTS population \citep{Glassgold_2007}. TTS, with their strong X-ray emission, effectively ionize neon in the surrounding disk atmospheres. X-rays penetrate deep into the disk and heat the gas, making them the dominant ionizing source for [Ne {\sc II}] emission in these stars. While EUV radiation also contributes to neon ionization, its role is limited by rapid absorption in the disk's upper layers, rendering X-rays more effective in TTS. Studies have shown some correlations between neon line luminosity and both mass accretion rate and X-ray luminosity, though these results are primarily limited to class II sources. Beyond this, no significant correlations have been observed between neon emission and other stellar parameters. The evidence for a direct correlation between X-ray luminosity ($L_X$) and neon emission remains inconclusive, with some studies suggesting a weak correlation while others find no significant trend. This inconsistency is likely due to the complex geometry of the disk and the influence of additional factors such as stellar jets. The region of [Ne {\sc II}] formation is typically within 20 AU of the star, as proposed by \citet{Glassgold_2007}, with emission potentially arising from various processes such as photoevaporative winds and shocks. However, the specific mechanisms driving neon ionization, particularly the relative contributions of X-rays and EUV radiation and the effects of disk and stellar properties, remain under investigation, with distinct challenges in understanding these processes in HAeBe stars.

In HAeBe stars, neon lines have been detected in nine sources in literature, though the underlying mechanisms remain largely unexplored. In this study, we examined the largest sample of HAeBe stars with neon detection. Notably, no correlation was found between both the normalised accretion luminosity and X-ray luminosity with [Ne {\sc II}] line luminosity. Our analysis suggests that EUV radiation is the primary driver of neon emission in our sample of HAeBe stars. This conclusion is based on the high detection rate of the [S {\sc III}] line, which is commonly found in EUV-dominated regions, and the strong positive correlation between the [Ne {\sc II}], [Fe {\sc II}], and [S {\sc III}] lines. Additionally, the high observed ratio of [Ne {\sc III}] to [Ne {\sc II}] emission lines (greater than 0.1) in sources where both lines are detected further supports this. Furthermore, considering the stellar properties and the mechanisms discussed, the observed neon emission likely originates from either EUV photoevaporation processes or irradiated disk atmospheres. The Meeus classification criteria and the outcomes reported by \cite{baldovin_2012} further support the predominance of photoevaporative winds as the emission mechanism. However, it is important to acknowledge that our sample lacked direct measurements of EUV emission. Therefore, pinpointing the exact processes definitively requires further investigation with high-resolution instruments, a larger sample size, and the inclusion of direct EUV observations. Additionally, a more comprehensive understanding of how specific disk configurations influence neon emission is crucial for a complete characterization of the underlying mechanisms. 

 High-resolution observations with JWST-MIRI may potentially reveal Neon emission in cases where low intrinsic line strength is buried by high continuum levels in Herbig stars. However, the detected values are unlikely to exceed the upper limits reported in this study. This consideration is critical for interpreting both the current findings and future investigations. Future work will therefore focus on expanding the sample size and acquiring data with advanced instruments, building upon the groundwork laid by this study. With the recent advancements in detecting forbidden neon lines in protostars and YSOs using JWST \citep{Espaillat_2023,Narang_2024,2024_Tychoniec,2024Sellek}, this catalog holds great potential to greatly advance our understanding of high-energy phenomena associated with HAeBe stars.
\section{Conclusions}
\label{sec:conclusions}
The paper presents an analysis of Spitzer MIR spectra from 78 well-known HAeBe stars, 25 of which exhibited distinct neon emission. The study aims to understand the emission region and associated high-energy processes contributing to line formation. We draw the following conclusions based on these data:
\begin{itemize}
    \item The majority of Ne-emitting sources have early spectral types ($T_{eff}$ $>$15000K). However, the uniform distribution of  $L_{[Ne \;{\sc II}]}$/$L_*$ across spectral types suggests that neon emission is not directly linked to spectral type. No clear distinctions in other stellar parameters were found between sources with and without neon emission.
    \item The lack of correlation between $L_{acc}$/$L_*$ and $L_{[Ne \;{\sc II}]}$/$L_*$ suggests that the neon line formation in our sample may not be directly driven by the accretion process.
    \item The formation of neon lines in our sample seems to be independent of X-ray irradiation, as evidenced by the absence of correlation between $L_{X}$/$L_*$ and $L_{[Ne \;{\sc II}]}$/$L_*$.
    \item Given the Meeus classification criteria and disk morphology, it is possible that the neon emission lines in these stars are related to photoevaporative winds and irradiated disk atmospheres.
    \item The moderate positive correlation between [Ne {\sc II}] and [O {\sc I}] indicates a shared emission region. However, since [Ne {\sc II}] traces a broader range of physical environments compared to [O {\sc I}] and considering the statistical uncertainty due to the small sample size, a larger dataset and higher-resolution spectra are needed to more precisely localize the emission region.
    \item The strong positive correlation between [Ne {\sc II}], [Fe {\sc II}], and [S {\sc III}] lines, suggests that neon emission lines in HAeBe stars are prevalent in regions with high-energy EUV radiation. Jets and outflows are also possible excitation regions for neon, warranting further investigation in HAeBe stars.
    \item The [Ne {\sc III}]-to-[Ne {\sc II}] line flux ratio for the 8 sources with both lines detected ranged mostly between 0.1 and 1, suggesting EUV as the main cause of  Ne line emission.
\end{itemize}
In summary, this study (to the best of our knowledge) represents the most extensive compilation of HAeBe stars exhibiting forbidden neon lines within the MIR spectrum, featuring numerous initial detections of  Ne lines. Photoevaporative winds or irradiated disk atmospheres are likely the source of neon emission in our sample of HAeBe stars. Furthermore, the high [Ne {\sc III}]-to-[Ne {\sc II}] line flux ratio ($>$0.1) in seven sources, coupled with their correlation with lines tracing EUV-irradiated regions, suggests that EUV radiation is the dominant high-energy photon source. However, to constrain the region of emission, and definitive mechanism of [Ne {\sc II}] emission, observations with a more sensitive instrument and a significantly larger sample are crucial.

\section*{Acknowledgements}
The authors thank the anonymous referee for the constructive report which has helped improve the overall quality of the paper. We would like to thank the Science \& Engineering Research Board (SERB), a statutory body of Department of Science \& Technology (DST), Government of India, for funding our research under grant number CRG/2023/005271. We are grateful to the Centre for Research, CHRIST (Deemed to be University), Bangalore, for the research grant extended to carry out the current project through the SEED money project (SMSS-2335, 11/2023). This work is based [in part] on observations made with the Spitzer Space Telescope, operated by the Jet Propulsion Laboratory, California Institute of Technology, under a contract with NASA. The X-Shooter data used in this research is based on observations collected at the European Southern Observatory under ESO programmes 0101.C-0866, 0101.C-0902, 084.C-0952, 084.C-0952A, 084.C-1095, 085.B-0751, 088.C-0218, 090.D-0212, 091.C-0934, 093.D-0415, and 094.C-0233. This research also utilizes data from eROSITA, the soft X-ray instrument aboard SRG, a joint Russian-German science mission supported by the Russian Space Agency (Roskosmos) in collaboration with the Russian Academy of Sciences' Space Research Institute (IKI) and the Deutsches Zentrum für Luft- und Raumfahrt (DLR). The SRG spacecraft was built by Lavochkin Association (NPOL) and its subcontractors and is operated by NPOL with support from the Max Planck Institute for Extraterrestrial Physics (MPE). The development and construction of the eROSITA X-ray instrument was led by MPE, with contributions from the Dr. Karl Remeis Observatory Bamberg \& ECAP (FAU Erlangen-Nuernberg), the University of Hamburg Observatory, the Leibniz Institute for Astrophysics Potsdam (AIP), and the Institute for Astronomy and Astrophysics of the University of Tübingen, supported by DLR and the Max Planck Society. The Argelander Institute for Astronomy at the University of Bonn and the Ludwig Maximilians Universität Munich also participated in the science preparation for eROSITA. This research has also made use of data obtained from the 4XMM XMM-Newton serendipitous source catalogue compiled by the XMM-Newton Survey Science Centre. Additionally, This work presents results from the European Space Agency (ESA) space mission Gaia. Gaia data are being processed by the Gaia Data Processing and Analysis Consortium (DPAC). Funding for the DPAC is provided by national institutions, in particular the institutions participating in the Gaia MultiLateral Agreement (MLA). The Gaia mission website is \url{https://www.cosmos.esa.int/gaia}. The Gaia archive website is \url{https://archives.esac.esa.int/gaia}. This publication also makes use of data products from the Wide-field Infrared Survey Explorer (WISE), a joint project of the University of California, Los Angeles, and the Jet Propulsion Laboratory/California Institute of Technology, funded by the National Aeronautics and Space Administration (NASA). Additionally, this work incorporates data from the Two Micron All Sky Survey (2MASS), a joint project of the University of Massachusetts and the Infrared Processing and Analysis Center/California Institute of Technology, funded by NASA and the National Science Foundation (NSF). This publication also makes use of VOSA, developed under the Spanish Virtual Observatory (https://svo.cab.inta-csic.es) project funded by MCIN/AEI/10.13039/501100011033/ through grant PID2020-112949GB-I00. VOSA has been partially updated by using funding from the European Union's Horizon 2020 Research and Innovation Programme, under Grant Agreement nº 776403 (EXOPLANETS-A). We also acknowledge the SIMBAD database and the VizieR online library service for their assistance in the literature survey.

%%%%%%%%%%%%%%%%%%%%%%%%%%%%%%%%%%%%%%%%%%%%%%%%%%
\section*{Data Availability}

The data utilized in this study can be accessed through the following archives and catalogs: the X-Shooter, the Gaia Archive, the Two Micron All-Sky Survey (2MASS) Catalog, the AllWISE Catalog, the XMM-Newton Data Archive, the eROSITA-DE Data Release 1 (eRODat) Archive, and the Chandra Data Archive. The data from the Combined Atlas of Sources with Spitzer IRS Spectra (CASSIS) can be accessed at \url{https://cassis.sirtf.com/}. The complete results produced by this study will be made available to interested parties upon request to the corresponding author.



\bibliographystyle{mnras}
\bibliography{example} % if your bibtex file is called example.bib

%\begin{thebibliography}{99}
%\end{thebibliography}

%%%%%%%%%%%%%%%%%%%%%%%%%%%%%%%%%%%%%%%%%%%%%%%%%%

%%%%%%%%%%%%%%%%% APPENDICES %%%%%%%%%%%%%%%%%%%%%

\appendix
\appendix

\section{Neon line detection}
We analyzed 78 continuum-subtracted, dereddened spectra using polynomial fitting as explained in Section \ref{sec: detection}. Noise levels were determined by calculating the standard deviation ($\sigma$) in regions around the expected line centers. Lines were considered detected if their peak flux exceeded a 3$\sigma$ threshold. We identified forbidden neon emission lines in 25 sources. For these sources, we fitted Gaussian models to the [Ne {\sc II}] and [Ne {\sc III}] lines using the \texttt{fit\_lines} function from \textsc{Specutils} \citep{nicholas_earl_2023_10016569}. Figures \ref{fig:neii}, \ref{fig:neii_1}, and \ref{fig:neiii} show the spectra for sources with [Ne {\sc II}] and [Ne {\sc III}] detections, covering a wavelength range of 12.56 $\micron$ -- 13.16 $\micron$ and 15.3 $\micron$ -- 15.78 $\micron$, respectively. The fitted Gaussian is represented by a blue solid line, with $\sigma$ and 3$\sigma$ thresholds indicated by green and red dashed lines, respectively.
\begin{figure*}
         \centering
         \includegraphics[width=0.9\textwidth]{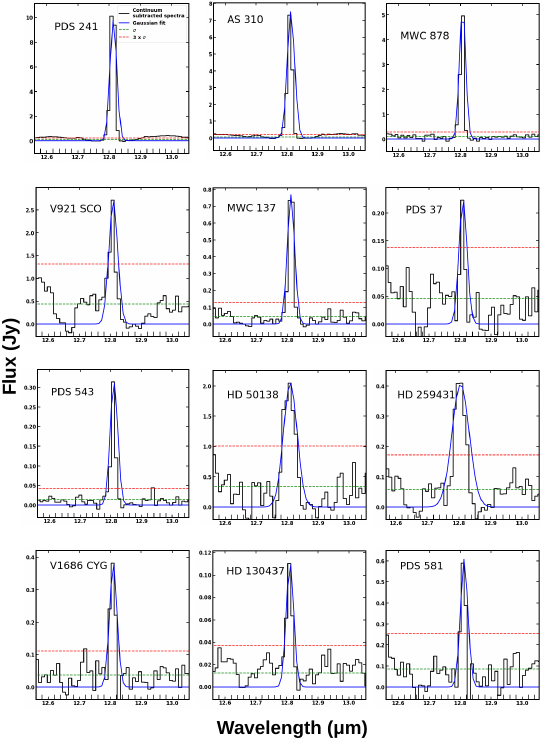}
          \caption{The MIR spectra of 24 HAeBe stars in our sample showing the presence of distinct [Ne {\sc II}] line. The fitted Gaussian is shown as blue solid line, while the $\sigma$ and 3$\sigma$ thresholds are marked by green and red dashed lines.}
          \label{fig:neii}
\end{figure*}
\begin{figure*}
         \centering
         \includegraphics[width=0.9\textwidth]{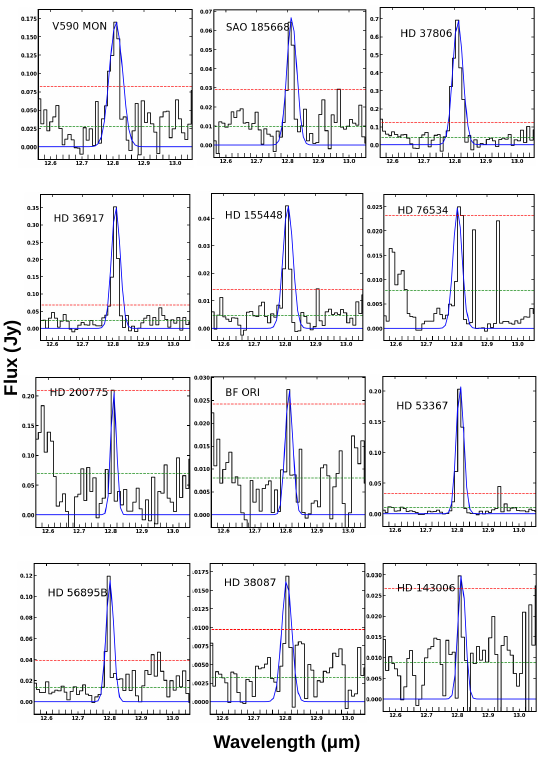}
          \caption{continuation.}
          \label{fig:neii_1}
\end{figure*}

 \begin{figure*}
         \centering
         \includegraphics[width=\textwidth]{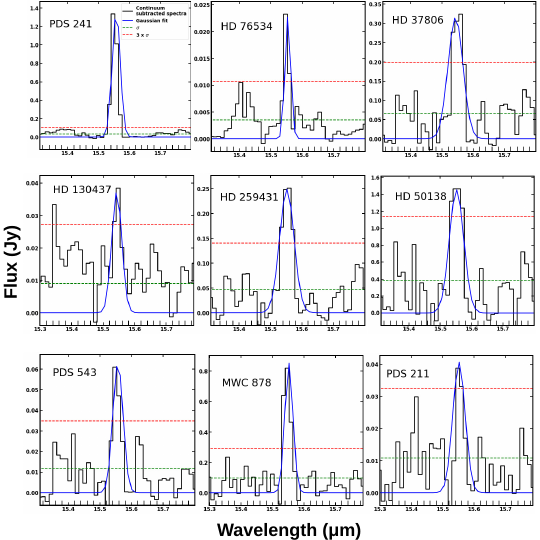}
          \caption{The MIR spectra of 9 HAeBe stars in our sample showing the presence of distinct [Ne {\sc III}] line. The fitted Gaussian is shown as blue solid line, while the $\sigma$ and 3$\sigma$ thresholds are marked by green and red dashed lines.}
          \label{fig:neiii}
\end{figure*}

\section{X-ray data reduction and analysis}
\label{appendix:X-ray data reduction and analysis}    
We retrieved eROSITA data from the eROSITA-DE Data Release 1 (eRODat) archive \citep{2024_merloni_erositadr1}, which included light curves, spectra, and response files from each of the seven telescope modules (TMs), as well as combined data from all the telescope modules. For this study, we utilized the combined spectral and response files from all seven TMs. %The spectra were grouped to ensure a minimum of 5 or more counts/bin. 
 For \textit{XMM-Newton}, we utilized archival data from the European Photon Imaging Camera (EPIC), specifically the PN \citep{2001Struder_pn} and MOS \citep{2001Turner_mos} detectors. The data reduction process was conducted using the XMM-Newton Science Analysis System (SAS) software, version 21.0 \citep{2017Gabriel_sas}. To generate the event files for the MOS and PN detectors, we employed the tasks \texttt{emproc} and \texttt{epproc}, respectively, while utilizing the latest available calibration files (CCF) during data reduction to ensure accurate results. The event files were checked for background flaring caused by high-energy particles, were corrected based on threshold rates set above the steady background levels. No pile-up was detected in our data. The EPIC spectra were obtained using the standard filtering criteria recommended by the XMM-Newton Science Operations Centre. We utilized the task \texttt{epicspeccombine} to merge the spectra from the three EPIC cameras (PN, MOS1, and MOS2) for all available observations. This process resulted in a single combined spectrum with a better signal-to-noise ratio along with its corresponding calibration matrices (rmf, arf) and background (bkg) files. In some cases (3 sources), only the MOS or PN spectra were available. We adopted the same spectral fitting procedure described in \citet{2024Anilkumar}, performing spectral analysis with \texttt{XSPEC version 12.13.1} \citep{1996Arnaud_xspec}. The X-ray spectra from eROSITA and XMM-Newton for all the sources in our sample were grouped to contain a minimum of 5 counts per bin. 
 
For the spectral fitting, we employed the “APEC” plasma emission model \citep{Smith_2001} and the “TBABS” absorption model, with the solar abundances set to \citet{1998Grevesse} and \citet{Wilms_2000}, respectively. Spectral models were initially fitted with a single-temperature (1T) component. A two-temperature (2T) model was applied only when necessary to achieve an optimal fit. The absorbing column density ($N_H$) was fixed based on A$_V$ using the equation from \citet{1996Ryter}, with the global metal abundance set to 0.3 of the solar value \citep{2003Imanishi, Getman2005}. \textcolor{black}{We employed $\mathrm{\chi}^2$-statistics for bright sources ($>$~100 counts), and grouped the spectra to ensure a minimum of 15 counts per bin, which conform to a Gaussian distribution \citep{bevington2003data}. The quality of the spectral fit was based on the null-hypothesis probability (P$_\mathrm{NULL}~>$ 5\%) for $\mathrm{\chi}^2$-statistics. For faint sources ($<$~100 counts), characterized by spectra with less than 10 counts per bin, c-statistics was adopted.} The unabsorbed flux in the 0.3 -- 8.0~keV range was obtained using the “CFLUX” model and luminosity was calculated. \textcolor{black}{Although the fit parameters derived from the model include temperature (kT) and emission measure (EM), we will report only the total X-ray luminosities of our targets, as these are the most critical parameters for our study.}
\section{ Ne flux and luminosity upper limits}
For the 53 sources where the line did not exceed the detection threshold, the flux upper limit was calculated using the instrumental FWHM, as detailed in Section \ref{sec: Ne line Flux and luminosity measurements}. This upper limit was then used to determine the corresponding upper limit on luminosity. The obtained flux and luminosity upper limits for [Ne {\sc II}] and [Ne {\sc III}] for these 53 sources, along with the $L_X$ and other parameters are detailed in the Table \ref{sec: luminosities 53}.
 \begin{table*}
    \centering
    \caption{The table represents the neon line flux ($F_{[Ne \;{\sc II}]}$, $F_{[Ne \;{\sc III}]}$) and luminosity upper limits ($L_{[Ne \;{\sc II}]}$, $L_{[Ne \;{\sc III}]}$) as well as the available $L_{X}$ of the 53 sources without neon detection. The distance (Dist.), log($L_*$), $A_V$ were obtained from \citet{Vioque_2018}, while the log($L_{acc}$) were from \citet{Guzman-Diaz_2021}. }
    \label{sec: luminosities 53}
    \begin{tabular}{ccccccccccc}
    \hline
        \textbf{Name} & \textbf{Dist.} & \textbf{log($L_*$)} & \textbf{$A_V$} & \textbf{log($L_{acc}$)} & \textbf{$F_{[Ne \;{\sc II}]}$} & \textbf{$L_{[Ne \;{\sc II}]}$} & \textbf{$F_{[Ne \;{\sc III}]}$} & \textbf{$L_{[Ne \;{\sc III}]}$} & \textbf{$L_{X}$}& \textbf{instrument} \\ 
        ~ & (pc) & ($L_{\odot}$) & (mag) & ($L_{\odot}$) &(10$^{-14}$erg s$^{-1}$cm$^{-2}$) & (10$^{30}$erg s$^{-1}$) &(10$^{-14}$erg s$^{-1}$cm$^{-2}$) & (10$^{30}$erg s$^{-1}$) &(10$^{30}$erg s$^{-1}$)& ~ \\ \hline
        V594 Cas & 569.2 & 2.13 & 1.9 & 1.58 & $<$4.01 & $<$1.56 & $<$3.56 & $<$1.38 & ... & ... \\ 
        HD 9672 & 57.1 & 1.17 & 0 & 0.37 &$<$ 0.53 & $<$0.002 & $<$0.40 & $<$0.002 & ... & ... \\ 
        HD 17081 & 106.7 & 2.58 & 0 & 1.8 & $<$0.51 & $<$0.01 & $<$0.51 & $<$0.01 & ... & ... \\ 
        BD+30 549 & 295.4 & 1.54 & 1.73 & 0.75 & $<$12.4 & $<$1.3 & $<$0.48 & $<$0.05 & 19.5$\pm$54.8 & Chandra \\ 
        V892 Tau & 117.5 & 0.13 & 4.87 & ... & $<$282 & $<$4.65 & $<$207 & $<$3.42 & 4.07 $\pm$ 0.563 & Chandra \\ 
        HD 35929 & 387.4 & 1.79 & 0 & 1.16 & $<$2.19 & $<$0.39 & $<$0.89 & $<$0.16 & ... & ... \\ 
        HD 36112 & 160.3 & 1.04 & 0.15 & 0.1 & $<$3.35 & $<$0.10 & $<$2.48 & $<$0.08 & 0.09$\pm$0.0132 & XMM-N \\ 
        HD 244604 & 420.6 & 1.46 & 0.14 & 0.71 & $<$1.66 & $<$0.35 & $<$1.18 & $<$0.25 & 0.15$\pm$0.09 & Chandra \\ 
        V380 Ori & 481.7 & 2 & 2.21 & 1.17 & $<$5.12 & $<$1.42 & $<$3.55 & $<$0.98 & 81.3$\pm$33.7 & Chandra \\ 
        HD 37258 & 362.7 & 1.24 & 0.06 & 0.58 & $<$2.1 & $<$0.33 & $<$1.18 & $<$0.19 & 2.24$\pm$0.24 & XMM-N \\ 
        HD 37357 & 649.6 & 2.04 & 0 & 1.13 & $<$2.52 & $<$1.27 & $<$1.42 & $<$0.72 & 0.46$\pm$0.08 & XMM-N \\ 
        RR Tau & 773.4 & 2.01 & 1.55 & 1.38 & $<$6.05 & $<$4.33 & $<$0.96 & $<$0.69 & ... & ... \\ 
        HD 38120 & 405 & 1.72 & 0.21 & 1.04 & $<$3.42 & $<$0.67 & $<$4.83 & $<$0.95 & 3.34$\pm$3.19 & eROSITA \\ 
        HD 39014 & 44.1 & 1.42 & 0 & 0.84 & $<$0.63 & $<$0.001 & $<$0.39 & $<$0.001 & 0.004 & eROSITA \\ 
        V1818 Ori & 695 & 2.96 & 3.717 & 1.79 & $<$8.07 & $<$4.66 & $<$8.25 & $<$4.77 & ... & ... \\ 
        HD 250550 & 697.1 & 1.94 & 0 & 1.39 & $<$2.47 & $<$1.44 & $<$2.27 & $<$1.32 & 1.41$\pm$0.976 & Chandra \\ 
        LkHa 215 & 713.1 & 2.57 & 2.02 & 1.9 & $<$9.71 & $<$5.91 & $<$0.62 & $<$0.37 & ... & ... \\ 
        HD 50083 & 1089.8 & 4.04 & 0.68 & 2.66 & $<$2.46 & $<$3.5 & $<$1.04 & ... & 338$\pm$142 & eROSITA \\ 
        NX Pup & 1672.5 & 2.46 & 0 & ... & $<$3.04 & $<$10.2 & $<$2.07 & $<$6.92 & ... & ... \\ 
        PDS 27 & 2552.6 & 4.15 & 5.03 & 2.59 & $<$9.47 & $<$73.8 & $<$11 & $<$85.8 & ... & ... \\ 
        HD 58647 & 318.5 & 2.44 & 0.37 & 1.68 & $<$3.62 & $<$0.44 & $<$2.05 & $<$0.25 & ... & ... \\ 
        HD 72106B & 597.2 & 1.85 & 0.50 & ... & $<$11 & $<$4.71 & $<$1.45 & $<$0.62 & ... & ... \\ 
        V388 Vel & 2466.9 & 2.45 & 3.99 & 1.32 & $<$12.2 & $<$88.6 & $<$1.97 & $<$14.3 & ... & ... \\ 
        HD 85567 & 1023 & 3.19 & 0.89 & ... & $<$12.7 & $<$16 & $<$2.87 & $<$3.59 & ... & ... \\ 
        HD 95881 & 1168.3 & 2.85 & 0 & 1.97 & $<$4.08 & $<$6.67 & $<$2.88 & $<$4.7 & ... & ... \\ 
        HD 97048 & 184.8 & 1.54 & 0.9 & 0.99 & $<$57.6 & $<$2.35 & $<$5.36 & $<$0.22 & 0.12$\pm$0.04 & XMM-N \\ 
        HD 98922 & 688.8 & 3.03 & 0.09 & 2.09 & $<$22.2 & $<$12.6 & $<$13.7 & $<$7.77 & ... & ... \\ 
        HD 101412 & 411.3 & 1.58 & 0.21 & 0.87 & $<$1.82 & $<$0.37 & $<$1.61 & $<$0.33 & 5.56$\pm$1.28 & eROSITA \\ 
        HD 104237 & 108.4 & 1.33 & 0 & 0.46 & $<$10 & $<$0.14 & $<$5.87 & $<$0.08 & 8.13$\pm$3.18 & Chandra \\ 
        Hen 2-80 & 753.5 & 2.12 & 2.97 & 2.86 & $<$34.9 & $<$23.7 & $<$3.26 & $<$2.21 & ... & ... \\ 
        DK Cha & 242.9 & 0.47 & 8.12 & ... & $<$9.99 & $<$0.70 & $<$7.46 & $<$0.53 & 16.6$\pm$3.9 & XMM-N \\ 
        Hen 3-847 & 784.8 & 2.07 & 0.57 & ... & $<$21.8 & $<$16.1 & $<$20.1 & $<$14.8 & ... & ... \\ 
        PDS 69 & 642.5 & 2.7 & 1.6 & 1.99 & $<$2.39 & $<$1.18 & $<$1.15 & $<$0.57 & 1.58$\pm$0.36 & Chandra \\ 
        DG Cir & 832.9 & 1.58 & 3.94 & 1.32 & $<$1.4 & $<$1.16 & $<$1.5 & $<$1.24 & ... & ... \\ 
        HD 132947 & 381.6 & 1.61 & 0 & 0.98 & $<$1.04 & $<$0.18 & $<$0.43 & $<$0.07 & ... & ... \\ 
        HD 135344B & 135.8 & 0.79 & 0.23 & -0.14 & $<$0.94 & $<$0.02 & $<$0.60 & $<$0.01 & 0.35$\pm$0.02 & Chandra \\ 
        PDS 144S & 149.6 & -0.67 & 0.57 & ... & $<$2.11 & $<$0.06 & $<$1.17 & $<$0.03 & ... & ... \\ 
        HD 142527 & 157.3 & 0.96 & 0 & 0.52 & $<$4.13 & $<$0.12 & $<$3.6 & $<$0.12 & 0.18$\pm$0.024 & XMM-N \\ 
        HR 5999 & 161.1 & 1.72 & 0.33 & 1.16 & $<$7.25 & $<$0.22 & $<$5.54 & $<$0.17 & 0.08$\pm$0.04 & Chandra \\ 
        PDS 415N & 144.2 & 0.44 & 1.48 & ... & $<$2.21 & $<$0.05 & $<$2.72 & $<$0.07 & 3.74$\pm$0.85 & eROSITA \\ 
        Hen 3-1191 & 1661.5 & 3.49 & 3.84 & 2.63 & $<$21.3 & $<$70.2 & $<$7.9 & $<$26.1 & ... & ... \\ 
        HD 149914 & 158.8 & 2.09 & 0.95 & 1.24 & $<$0.40 & $<$0.01 & $<$0.43 & $<$0.01 & ... & ... \\ 
        HD 150193 & 150.8 & 1.37 & 1.55 & 0.53 & $<$5.76 & $<$0.16 & $<$5.39 & $<$0.13 & 0.29$\pm$0.18 & Chandra \\ 
        KK Oph & 221.1 & 0.71 & 2.7 & ... & $<$5.6 & $<$0.33 & $<$4.26 & $<$0.25 & ... & ... \\ 
        HD 163296 & 101.5 & 1.2 & 0 & 0.36 & $<$8.07 & $<$0.09 & $<$6.43 & $<$0.08 & 0.78$\pm$0.25 & Chandra \\ 
        VV Ser & 419.7 & 1.95 & 2.91 & 1.51 & $<$2.42 & $<$0.51 & $<$1.8 & $<$0.38 & ... & ... \\ 
        HD 179218 & 266 & 2.05 & 0.53 & 1.21 & $<$8.48 & $<$0.72 & $<$14.6 & $<$1.24 & ... & ... \\ 
        WW Vul & 503.5 & 1.42 & 0.95 & 0.58 & $<$0.73 & $<$0.22 & $<$0.66 & $<$0.20 & ... & ... \\ 
        MWC 623 & 3279.8 & 4.58 & 3.77 & ... & $<$9.45 & $<$122 & $<$4.02 & $<$51.8 & ... & ... \\ 
        V1295 Aql & 870.9 & 2.9 & 0.40 & 1.92 & $<$3.83 & $<$3.48 & $<$2.31 & $<$2.1 & 0.72$\pm$0.57 & Chandra \\ 
        V373 Cep & 922.1 & 2.29 & 3.07 & 1.62 & $<$22.3 & $<$22.6 & $<$4.07 & $<$4.14 & ... & ... \\ 
        V669 Cep & 977.6 & 2.6 & 3.05 & ... & $<$5.35 & $<$6.12 & $<$4.82 & $<$5.51 & 58.4$\pm$20 & XMM-N \\ 
        MWC 657 & 3164.2 & 4.62 & 5.03 & ... & $<$4.45 & $<$53.3 & $<$2.65 & $<$31.7 & ... & ... \\ \hline
    \end{tabular}
\end{table*}

%%%%%%%%%%%%%%%%%%%%%%%%%%%%%%%%%%%%%%%%%%%%%%%%%%

% Don't change these lines
\bsp	% typesetting comment
\label{lastpage}
\end{document}